\pdfoutput=1
\newcommand*{\ATLASLATEXPATH}{latex/}
\documentclass[cernpreprint,texlive=2016,txfonts,UKenglish]{latex/atlasdoc}

\usepackage[subcaption=true]{\ATLASLATEXPATH atlaspackage}
\usepackage{\ATLASLATEXPATH atlasbiblatex}

\usepackage{\ATLASLATEXPATH atlasphysics}

\graphicspath{{logos/}{figures/}}

\usepackage{multirow}
\usepackage{vbfz13paper-defs}
\usepackage{enumitem}
\usepackage[utf8]{inputenc}
\usepackage{arydshln}

\AtlasTitle{Measurement of the cross-section for electroweak production of dijets in association with a $Z$ boson in $pp$ collisions at $\sqrt{s}=13$~TeV with the ATLAS detector}

\author{The ATLAS Collaboration}

\usepackage{authblk}
\renewcommand\Authands{,\xspace} 
\renewcommand\Affilfont{\itshape\small} 

\AtlasRefCode{STDM-2016-09}

\PreprintIdNumber{CERN-EP-2017-115}

 \AtlasJournal{Phys.\ Lett.\ B. 775 (2017) 206}
 \AtlasDOI{10.1016/j.physletb.2017.10.040}

\AtlasAbstract{
    The cross-section  for the production of two 
    jets in association with a leptonically decaying \Z boson (\zjj) is measured in proton--proton collisions
 at a centre-of-mass energy of 13~TeV, using data recorded with the ATLAS detector at the Large Hadron Collider, corresponding to an integrated luminosity of 3.2~\ifb.
    The electroweak \zjj  cross-section is extracted in a fiducial region chosen to enhance the electroweak contribution relative to the
    dominant Drell--Yan \zjj process, which is constrained
    using a data-driven approach. The measured fiducial electroweak cross-section is \sigmaewk~=~\ewkxsec~fb for dijet invariant mass greater than 250~GeV, and 
    $\ewkxsecHM$~fb for dijet invariant mass greater than 1~TeV. Standard Model predictions are in agreement with the measurements.
 The inclusive \zjj cross-section is also measured in six different fiducial regions with varying contributions from electroweak and Drell--Yan \zjj production.
}

\hypersetup{pdftitle={ATLAS document},pdfauthor={The ATLAS Collaboration}}

\begin{document}

\maketitle
 
\section{Introduction} 

\begin{figure}[htbp]
  \centering
  \begin{subfigure}[b]{0.45\textwidth}
    \includegraphics[scale=0.33]{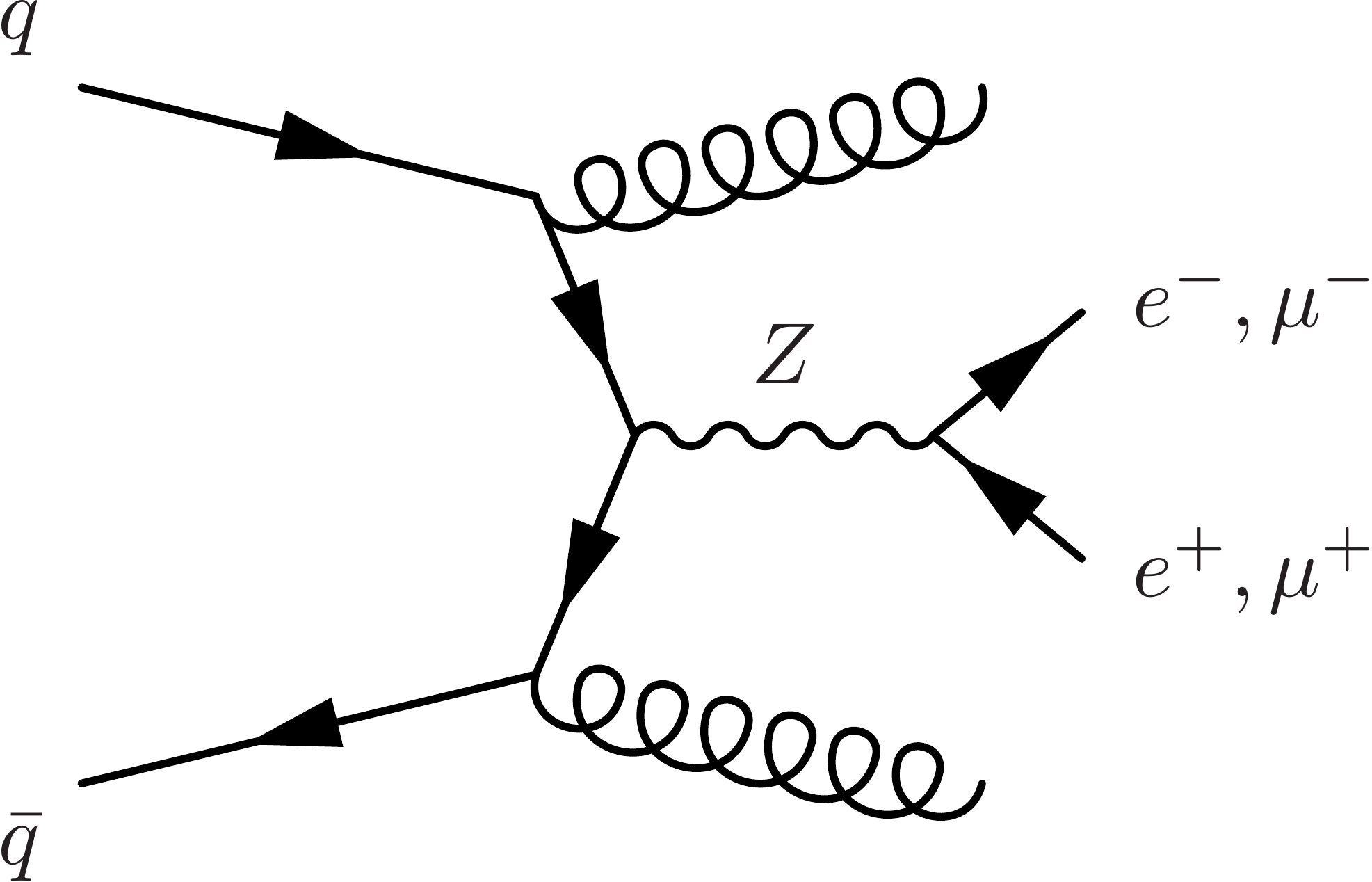}
    \caption{\dyzjj.}\label{fig:dyzjj}
  \end{subfigure}
  \hspace{0.5cm}
  \begin{subfigure}[b]{0.45\textwidth}
    \includegraphics[scale=0.33]{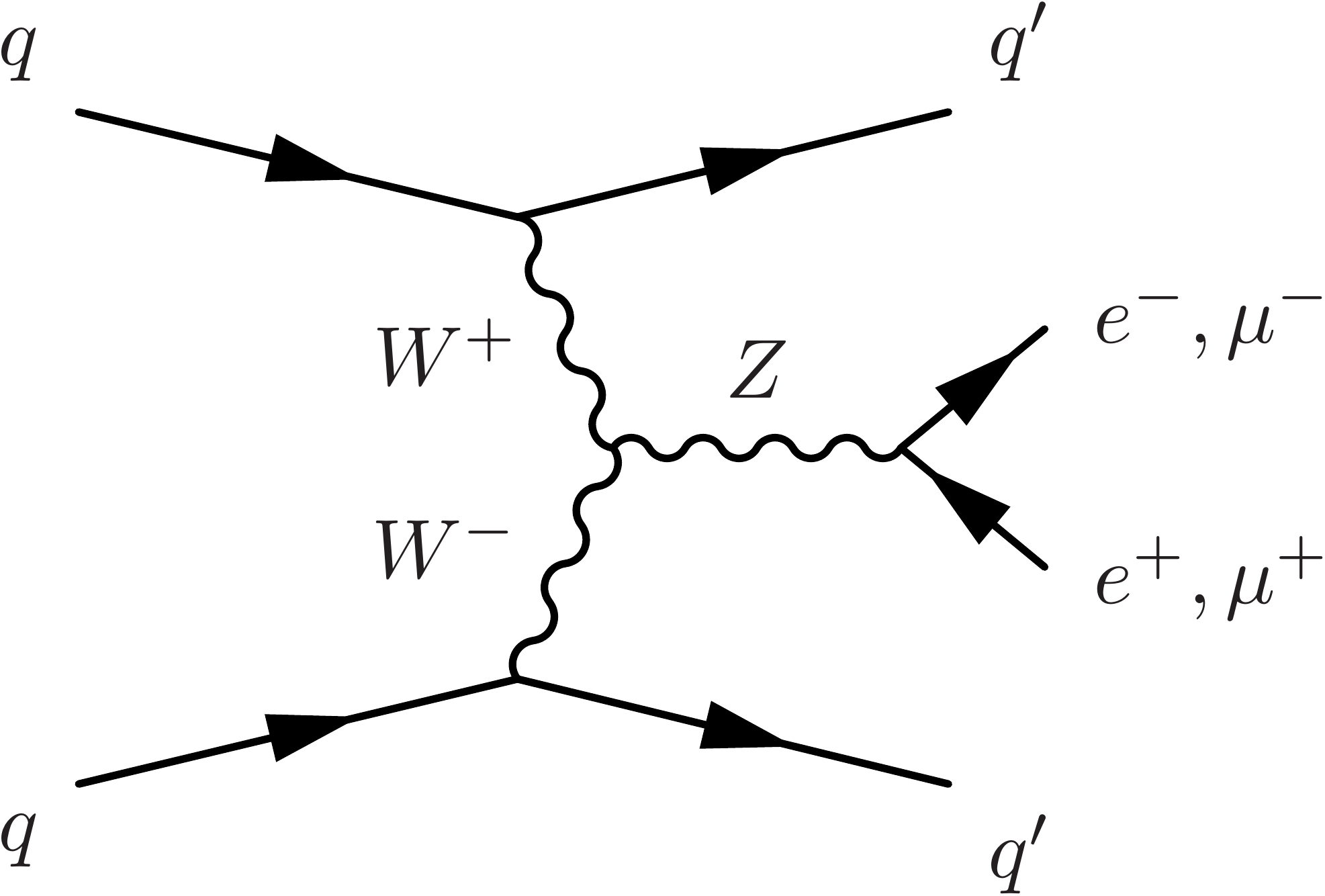} 
    \caption{\ewkzjj.}\label{fig:ewkzjj}
  \end{subfigure}
  \caption{Examples of leading-order Feynman diagrams for the two production mechanisms for a leptonically decaying \Z boson and at least two jets (\zjj) in proton--proton collisions: (a) QCD radiation from the incoming partons (\dyzjj) and (b) $t$-channel exchange of an EW gauge boson (\ewkzjj). }
  \label{fig:Zjj}
\end{figure}

At the Large Hadron Collider (LHC) 
events containing a \Z boson and at least two jets (\zjj) 
are produced predominantly via initial-state QCD radiation from the incoming partons in the Drell--Yan process (\dyzjj), as shown in Figure~\ref{fig:dyzjj}. 
In contrast, the production of \zjj events via $t$-channel electroweak gauge boson exchange
(\ewkzjj events), including the vector-boson fusion (VBF) process shown in Figure~\ref{fig:ewkzjj}, is a much rarer process. 
Such VBF processes for vector-boson production are of great interest as a 
`standard candle' for other VBF processes at the LHC: e.g., the production of Higgs bosons or the search for weakly interacting particles beyond the Standard Model. 

The kinematic properties of \zjj events allow some discrimination between the QCD and EW production mechanisms.
The emission of a virtual $W$ boson from the quark 
in \ewkzjj events results in the
presence of 
two high-energy jets, with moderate transverse momentum (\pT), 
 separated by a large interval in
rapidity ($y$)\footnote{ATLAS uses a right-handed coordinate system with its origin at the nominal
interaction point in the centre of the detector and the $z$-axis along the beam pipe. In the transverse plane, the $x$-axis
points from the interaction point to the centre of the LHC ring, the $y$-axis points upward, and
$\phi$ is the azimuthal angle
around the $z$-axis. The pseudorapidity is defined in terms of the polar angle $\theta$ as $\eta=-\ln\tan(\theta/2)$.
The rapidity is defined as $y=0.5\ln [(E+p_z)/(E-p_z)]$, where $E$ and $p_z$ are the energy and
longitudinal momentum respectively.
An angular separation between two objects is defined as $\Delta R = \sqrt{(\Delta \phi)^{2} + (\Delta \eta) ^{2}}$, where $\Delta\phi$ and $\Delta\eta$ are the separations in $\phi$ and $\eta$ respectively.
Momentum in the transverse plane is denoted by $\pt$.} 
and therefore with large dijet mass (\mjj) that characterises the \ewkzjj signal. 
A consequence of the exchange of a vector boson in Figure~\ref{fig:ewkzjj} is that there is no colour connection between the 
hadronic systems produced by the break-up of the two incoming protons.
As a result, \ewkzjj events are less likely to contain additional hadronic activity in the rapidity interval between the two high-\pT jets than corresponding \dyzjj events. 

The first observation of the \ewkzjj process and a measurement of the corresponding fiducial cross-section was performed by the ATLAS Collaboration
in \emph{pp} collisions at a centre-of-mass energy ($\sqrt{s}$) of $\SI{8}{\TeV}$\,\cite{8TeVVBFNote}.
The measurement is in agreement with predictions from the \POWHEG-\textsc{box} event generator\,\cite{powheg1,powheg2,powheg3} and allowed limits to
be placed on anomalous triple gauge couplings.
The cross-section for \ewkzjj production at $\sqrt{s}=\SI{8}{\TeV}$ has also been measured by the CMS Collaboration\,\cite{CMS8TeV}.
This Letter presents measurements of the cross-section for  \ewkzjj production and inclusive \zjj production at high dijet invariant mass in \emph{pp} collisions at $\sqrt{s}=\SI{13}{\TeV}$ 
using data corresponding to an integrated luminosity of \SI{3.2}{fb^{-1}} 
collected by the ATLAS detector at the LHC.
These measurements allow  the dependence of the cross-section on $\sqrt{s}$ to be studied.
The increased $\sqrt{s}$ allows exploration of higher dijet masses, where the \ewkzjj contribution to the total \zjj rate becomes more pronounced.

\section{ATLAS detector}

The ATLAS detector is described in detail in Refs.\,\cite{PERF-2007-01,ATLAS-TDR-19}.
It consists of an inner detector for tracking, surrounded by a thin superconducting solenoid, electromagnetic and hadronic calorimeters,
and a muon spectrometer incorporating three large superconducting toroidal magnet systems.
The inner detector is immersed in a 2\,T axial magnetic field and provides charged-particle tracking in the range $|\eta|< 2.5$. 

The calorimeters cover the pseudorapidity range $|\eta|< 4.9$. 
Electromagnetic calorimetry is provided by barrel and end-cap lead/liquid-argon (LAr) calorimeters in the region $|\eta|< 3.2$.
Within $|\eta| < 2.47$ the calorimeter 
is finely segmented in the lateral direction of the showers, allowing measurement of the energy and position of electrons,
and providing electron identification in conjunction with the inner detector.
Hadronic calorimetry is provided by the steel/scintillator-tile calorimeter, segmented into three barrel structures within $|\eta| < 1.7$, and two
hadronic end-cap calorimeters.
A copper/LAr hadronic calorimeter covers the $1.5 < |\eta| < 3.2$ region, and a forward copper/tungsten/LAr calorimeter with electromagnetic-shower identification capabilities covers the $3.1 < |\eta| < 4.9$ region.

The muon spectrometer comprises separate trigger and high-precision tracking chambers.
The tracking chambers cover the region $|\eta| < 2.7$ with
three layers of monitored drift tubes, complemented by cathode strip chambers in part of the forward region, where the hit rate is highest.
The muon trigger system covers the range $|\eta| < 2.4$ with resistive plate chambers in the barrel region, and thin gap chambers
in the end-cap regions. 

A two-level trigger system is used to select events of interest\,\cite{Aaboud:2016leb}.
The Level-1 trigger is implemented in hardware and uses a subset of the detector information to reduce the event rate to 
 around 100~kHz.
This is followed by the software-based high-level trigger system which reduces the event rate to about 1~kHz.

\section{Monte Carlo samples} 
\label{samples}

The production of \ewkzjj events was simulated at next-to-leading-order (NLO) accuracy in perturbative QCD using the 
\POWHEG-\textsc{box} v1 Monte Carlo (MC) event generator\,\cite{schissler_parton_2013,powheg2,powheg3} and, alternatively, 
at leading-order (LO) accuracy in perturbative QCD using the \SHERPA 2.2.0 event generator\,\cite{Gleisberg:2008ta}.
For modelling of the parton shower, fragmentation, hadronisation and underlying event (UEPS), 
\POWHEG-\textsc{box} was interfaced to \PYTHIA8\,\cite{Sjostrand:2014zea} with a dedicated set of parton-shower-generator parameters (tune) denoted \textsc{AZNLO}\,\cite{Aad:2014xaa} and the \textsc{CT10 NLO} parton distribution function (PDF) set\,\cite{Lai:2010vv}.
The renormalisation and factorisation scales were set to the $Z$ boson mass.
\SHERPA predictions used the \textsc{Comix}\,\cite{Gleisberg:2008fv} and \textsc{OpenLoops}\,\cite{Cascioli:2011va} matrix element event generators, and 
the CKKW method was used to combine the various final-state topologies from the matrix element and match them to the parton shower\,\cite{Catani:2001cc}.
The matrix elements were merged with the \SHERPA parton shower\,\cite{Schumann:2007mg} using the \textsc{ME+PS@LO} prescription\,\cite{Gehrmann:2012yg,Hoeche:2012yf}, and using \SHERPA's native dynamical scale-setting
algorithm to set the renormalisation and factorisation scales.
\SHERPA predictions used the \textsc{NNPDF30NNLO} PDF set\,\cite{Ball:2014uwa}.

The production of \dyzjj events was simulated using three event generators,
\SHERPA 2.2.1, \ALPGEN 2.14\,\cite{Mangano:2002ea} and \MGMCatNLO 2.2.2\,\cite{Alwall:2014hca}. 
\SHERPA provides $Z+n$-parton predictions calculated for up to two partons at NLO accuracy and up to four partons at LO accuracy in perturbative QCD.
\SHERPA predictions used the \textsc{NNPDF30NLO} PDF set together with the 
tuning of the UEPS parameters
developed by the \SHERPA authors using the \textsc{ME+PS@NLO} prescription. 
\ALPGEN is an LO event generator which uses explicit matrix elements for up to five partons and was interfaced to \PYTHIA 6.426\,\cite{Sjostrand:2006za} using the Perugia2011C tune\,\cite{perugia} and the CTEQ6L1 PDF set\,\cite{Pumplin:2002vw}.
Only matrix elements for light-flavour production in \ALPGEN are included, with heavy-flavour contributions modelled by the parton shower. 
\MGMCatNLO 2.2.2 (\MADGRAPH) uses explicit matrix elements for up to four partons at LO, and was interfaced to \PYTHIA8 with the \textsc{A14} tune\,\cite{ATL-PHYS-PUB-2014-021} and using the \textsc{NNPDF23LO} PDF set\,\cite{Ball:2012cx}.
For reconstruction-level studies, total \Z boson production rates predicted by all three event generators used to produce \dyzjj predictions are normalised using 
the next-to-next-to-leading-order (NNLO) predictions calculated with the \textsc{FEWZ 3.1} program\,\cite{Anastasiou:2003ds,Gavin:2010az,Li:2012wna} using the \textsc{CT10 NNLO} PDF set\,\cite{Lai:2010vv}.
However, when comparing particle-level theoretical predictions to detector-corrected measurements, the normalisation of quoted predictions is provided 
by the event generator in question rather than an external NNLO prediction.

The production of a pair of EW vector bosons (diboson), where one decays leptonically and
the other hadronically, or where both decay leptonically and are produced in association with two or more jets,
through $WZ$ or $ZZ$ production with at least one \Z boson decaying to leptons,
was simulated separately  using \SHERPA 2.1.1 and the \textsc{CT10 NLO} PDF set.

The largest background to the selected \zjj samples arises from \ttbar and single-top ($Wt$) production. These were generated using
\POWHEG-\textsc{box} v2 and \PYTHIA 6.428 with the Perugia2012 tune\,\cite{perugia}, 
and normalised using the cross-section calculated at NNLO+NNLL (next-to-next-to-leading log) accuracy using the \textsc{Top++2.0} program\,\cite{Czakon:2011xx}.

All the above MC samples were fully simulated through the \GEANT4\,\cite{Agostinelli2003250} simulation of the ATLAS detector\,\cite{SOFT-2010-01}.
The effect of additional $pp$ interactions (pile-up) in the same or nearby bunch crossings was also simulated, using \PYTHIA v8.186 with the \textsc{A2} tune\,\cite{ATL-PHYS-PUB-2012-003} and the \textsc{MSTW2008LO} PDF set\,\cite{Martin:2009iq}. 
The MC samples were reweighted so that the distribution of the average number of pile-up interactions per bunch crossing matches that observed in data. 
For the data considered in this Letter, the average number of interactions is 13.7.

\section{Event preselection}
\label{selection}

The \Z bosons are measured in their dielectron and dimuon decay modes.
Candidate events are selected using triggers requiring at least one identified electron or muon with transverse momentum thresholds of $\pT = 24$~GeV and 20~GeV respectively,
 with additional isolation requirements imposed
in these triggers. At higher transverse momenta, 
the efficiency of selecting candidate events
is improved through the 
use of additional electron and muon triggers without isolation requirements and with thresholds of $\pT = 60$~GeV and 50~GeV respectively.

Candidate electrons are reconstructed from clusters of energy in the electromagnetic
calorimeter matched to inner-detector tracks\,\cite{ATLAS-CONF-2016-024}. They must satisfy the {\em Medium} identification requirements described in Ref.~\cite{ATLAS-CONF-2016-024}
and have \pT~$>$~\SI{25}{GeV} and \abseta~$<$~2.47, excluding the transition region between the barrel and end-cap
calorimeters at 1.37~$<$~\abseta~$<$~1.52.
Candidate muons are identified as tracks in the inner detector matched and combined
with track segments in the muon spectrometer. They must satisfy the {\em Medium} identification requirements described in Ref.\,\cite{PERF-2015-10}, 
and have \pT~$>$~\SI{25}{GeV} and \abseta~$<$~2.4.
Candidate leptons must also satisfy a set of isolation criteria based on reconstructed tracks and calorimeter activity.
Events are required to contain exactly two leptons of the same flavour but of opposite charge.
The dilepton invariant mass must satisfy $ 81< \text{m}_{\ell \ell} <101 $ GeV.

Candidate hadronic jets are required to satisfy \pT~$>$~\SI{25}{GeV} and $|y|$~$<$~4.4.
They are reconstructed from clusters of energy in the calorimeter\,\cite{Lampl:1099735} using the
anti-$k_{t}$ algorithm\,\cite{Cacciari:2008gp,Cacciari:2011ma} with radius parameter \emph{R}~=~0.4.
Jet energies are calibrated by applying $\pT$- and $y$-dependent corrections derived from Monte Carlo simulation with additional in situ correction factors determined from data\,\cite{PERF-2016-04}.
To reduce the impact of pile-up contributions, all jets with $|y|<2.4$ and $\pt<60$~GeV are required to be compatible with having originated from the
primary vertex (the vertex with the highest sum of track $p^2_\mathrm{T}$), as defined by the jet vertex tagger algorithm\,\cite{ATLAS-CONF-2014-018}. 
Selected electrons and muons are discarded if they lie within $\Delta R=0.4$ of a reconstructed jet.
This requirement is imposed to remove non-prompt non-isolated leptons produced in heavy-flavour decays or from the decay in flight of a kaon or pion.

\section{Measurement of inclusive \zjj fiducial cross-sections} 
\label{inclusive}

\subsection{Definition of particle-level cross-sections} 

Cross-sections are measured for inclusive \zjj production that includes the \ewkzjj and \dyzjj processes, as well as diboson events.
The particle-level production cross-section for inclusive \zjj production in a given fiducial region $f$ is given by
\begin{equation}
  \sigma^{f} = \frac{N^f_\mathrm{obs} - N^f_\mathrm{bkg}}{L \cdot \mathcal{C}^f},
  \label{eqn:incl}
\end{equation}
where $N^f_\mathrm{obs}$ is the number of events observed in the data passing the selection requirements of the fiducial region under study at detector level, $N^f_\mathrm{bkg}$ is the corresponding number of expected background (non-\zjj) events, 
$L$ is the integrated luminosity corresponding to the analysed data sample, and $\mathcal{C}^f$ is a correction factor applied to
the observed data yields, which accounts for experimental efficiency and detector resolution effects, and is derived from MC simulation with data-driven efficiency and energy/momentum scale corrections.
This correction factor is calculated as:
\begin{equation*}
  \mathcal{C}^f = \frac{N^f_{\mathrm{det}}}{N^f_\mathrm{particle}},
\end{equation*}
where $N^f_{\mathrm{det}}$ is the number of signal events that satisfy the fiducial selection criteria at detector level in the
MC simulation, and $N^f_\mathrm{particle}$ is the number of signal events that pass the equivalent selection but at particle level.
These correction factors have values between $0.63$ and $0.77$, depending on the fiducial region.

With the exception of background from multijet and \emph{W}+jets processes (henceforth referred to together simply as multijet processes), contributions to $N^f_\mathrm{bkg}$ 
are estimated using the Monte Carlo samples described in Section~\ref{samples}. 
Background from multijet events is estimated from the data by reversing requirements on lepton identification or isolation to derive a template for the contribution of jets mis-reconstructed as lepton candidates as a function of dilepton mass.
Non-multijet background is subtracted from the template using simulation. The normalisation is derived by fitting the nominal dilepton 
mass distribution in each fiducial region with the sum of the multijet template and a template comprising signal and background contributions determined from simulation. 
The multijet contribution is found to be less than 0.3\% in each fiducial region. The contribution from \emph{W}+jets processes was checked using MC simulation and found to be much smaller 
than the total multijet background as determined from data.

At particle level, only final-state particles with proper lifetime $c\tau>$~10~mm are considered.
Prompt leptons are dressed using the
four-momentum combination of an electron or muon and all photons (not originating from hadron decays) within a
cone of size $\dr = 0.1$ centred on the lepton.
These dressed leptons are required to satisfy $\pt > 25$ GeV and  $|\eta| < 2.47$.
Events are required to contain exactly two dressed leptons of the same flavour but of opposite charge, and the dilepton invariant mass must satisfy $81< \text{m}_{\ell \ell} < 101$ GeV.
Jets are reconstructed using the anti-$k_{t}$ algorithm with radius parameter \emph{R}~=~0.4.
Prompt leptons and the photons used to dress these leptons are not included in the particle-level jet reconstruction.
All remaining final-state particles are included in the particle-level jet clustering.
Prompt leptons with a separation $\Delta R_{j,\ell} < 0.4$ from any jet are rejected.

The cross-section measurements are performed in the six phase-space regions defined in Table~\ref{table:phasespacescuts}. 
These regions are chosen to have varying contributions from \ewkzjj and \dyzjj processes.

\subsection{Event selection}
\label{selection:fiducial}

Following Ref.\,\cite{8TeVVBFNote}, events are selected in six detector fiducial regions. 
As far as possible, these are defined with the same kinematic requirements as the six phase-space regions in which the cross-section is measured (Table~\ref{table:phasespacescuts}).
This  minimises systematic uncertainties in the modelling of the acceptance.

\begin{table}[htbp]
  \footnotesize
  \centering
  \bgroup
  \def\arraystretch{1.8}
  \begin{tabular}{c|c|c|c|c|c|c}
    \hline \hline
           & \multicolumn{6}{c}{Fiducial region} \\
\hline
    Object & Baseline & High-mass & High-$p_\textrm{T}$ & \search & \search, & \control  \\ [-1.5ex]
           &          &           &                     &         & $\mjj>1$~TeV & \\
    \hline
    Leptons & \multicolumn{6}{c}{ $|\eta| < $ 2.47, $\pT > $ 25 GeV, $\Delta R_{j,\ell} > 0.4$ } \\
    \hline
    \multirow{2}{*}{Dilepton pair} & \multicolumn{6}{c}{ 81 $ < m_{\ell \ell} < $ 101 GeV } \\
                                   \cline{2-7}
                                   & \multicolumn{3}{c|}{ --- } & \multicolumn{3}{c}{ $\pT^{\ell \ell} > $ 20 GeV } \\
    \hline
    \multirow{3}{*}{Jets} & \multicolumn{6}{c}{ $|y| < $ 4.4 } \\
                            \cline{2-7}
                            & \multicolumn{2}{c|}{ \pTo{j_1} $>$ 55 GeV } & \pTo{j_1} $>$ 85 GeV & \multicolumn{3}{c}{ \pTo{j_1} $>$ 55 GeV } \\
                            \cline{2-7}
                            & \multicolumn{2}{c|}{ \pTo{j_2} $>$ 45 GeV } & \pTo{j_2} $>$ 75 GeV  & \multicolumn{3}{c}{ \pTo{j_2} $>$ 45 GeV }  \\
    \hline
    Dijet system & --- & \twomass{j}{j} $>$ 1 TeV & ---  & \twomass{j}{j} $>$ 250 GeV & \twomass{j}{j} $>$ 1 TeV  & \twomass{j}{j} $>$ 250 GeV \\
    \hline
    Interval jets & \multicolumn{3}{c|}{ --- } & \multicolumn{2}{c|}{\ngap = 0} & \ngap $\geq$ 1 \\
    \hline
    \zjj system & \multicolumn{3}{c|}{ --- } & \multicolumn{2}{c|}{$p_{\text{T}}^{\text{balance}} < $ 0.15}  & $p_{\text{T}}^{\text{balance},3} <$ 0.15 \\
    \hline \hline
  \end{tabular}
  \egroup
  \caption{ Summary of the particle-level selection criteria defining the six fiducial regions
  (see text for details).}
  \label{table:phasespacescuts}
\end{table}

The \baseline fiducial region represents an inclusive selection of events containing a leptonically decaying \Z boson and at least
two jets with $\pt > 45$~\GeV, at least one of which satisfies $\pt > 55$~\GeV. The two highest-$\pt$ (leading and sub-leading) jets in a given event define the dijet system.
The \baseline region is dominated by \dyzjj events.
The requirement  of $81  < \mll <  101$~GeV  suppresses other sources of dilepton events, such as \ttbar and $\Z\rightarrow\tau\tau$, as well as
 the multijet background.

Because the energy scale of the dijet system is typically higher in events produced by the \ewkzjj process than
in those produced by the \dyzjj process,
two subsets of the \baseline region are defined which probe the \ewkzjj contribution in different ways: 
in the \highmass fiducial region a high value of the invariant mass of the dijet system ($\mjj>1$~TeV) is required,
and in the \highpT fiducial region 
the minimum \pT  of the leading and sub-leading jets is increased to 85~GeV and 75~GeV respectively.
The \ewkzjj process typically produces harder jet transverse momenta and results in a harder dijet invariant mass spectrum than the \dyzjj process.

Three additional fiducial regions allow the separate contributions from the \ewkzjj and \dyzjj processes to be measured.
The \search fiducial region is designed to enhance the  \ewkzjj contribution relative to that from \dyzjj, particularly at high \mjj.
The \search region is derived from the \baseline region requiring $\mjj > 250$~GeV, a dilepton transverse momentum
of $\pT^{\ell \ell} >$~20 GeV, and that the normalised transverse momentum balance between the two leptons and the two highest transverse momentum jets satisfy $\ptsc < 0.15$.
The latter quantity is given by
\begin{equation}\label{eq:ptsc}
\ptsc = \frac{ \left|\vec{p}^{\, \ell_1}_\mathrm{T} + \vec{p}^{\, \ell_2}_\mathrm{T} + \vec{p}^{\, j_1}_\mathrm{T} + \vec{p}^{\, j_2}_\mathrm{T}\right|}{\left|\vec{p}^{\, \ell_1}_\mathrm{T}\right| + \left|\vec{p}^{\, \ell_2}_\mathrm{T}\right| + \left|\vec{p}^{\, j_1}_\mathrm{T}\right| + \left|\vec{p}^{\, j_2}_\mathrm{T}\right|}, 
\end{equation}
where $\vec{p}^{\, i}_\mathrm{T}$ is the transverse momentum vector of object $i$, $\ell_1$ and $\ell_2$ label the two leptons that
define the \Z boson candidate, and $j_1$ and $j_2$ refer to the leading and sub-leading jets.
These requirements help remove events in which the jets arise from pile-up or multiple parton interactions. The requirement on
\ptsc also helps suppress events in which the \pT of one or more jets is badly measured
and it enhances the \ewkzjj contribution, where the lower probability of additional radiation causes the \Z boson and the dijet system to be well balanced. 
The \search region requires a veto\,\cite{Rainwater:1996ud} on any jets with $\pT>25$~GeV reconstructed within the rapidity interval bounded by the dijet system (\ngap = 0).
A second fiducial region, denoted \search ($m_{jj}>1$~TeV), has identical selection criteria, except for a raised \mjj threshold of 1~TeV which further enhances the \ewkzjj contribution to the total \zjj signal rate.

In contrast, the \control fiducial region is designed to suppress the  \ewkzjj contribution relative to \dyzjj by requiring at least
one jet with $\pT>25$~GeV to be reconstructed within the rapidity interval bounded by the dijet system (\ngap $\geq$ 1).
In the \control region, the definition of the normalised transverse momentum balance is modified from that given in
Eq.~(\ref{eq:ptsc}) to include in the calculation of the numerator and denominator the \pT of
the highest \pT jet  within the rapidity interval bounded by the dijet system (\ptscthree).
In all other respects, the kinematic requirements in the \search region and \control region are identical.

\subsection{Detector-level results}

In the baseline region, $30\, 686$ events are selected in the dielectron channel and $36\, 786$ events are selected in the dimuon channel.
The total observed yields are in agreement with the expected yields within statistical uncertainties
in each dilepton channel. 
The largest deviation across all fiducial regions is a $2\sigma$ (statistical) difference between the expected to observed ratio in the electron versus muon channel in the \highpT region.

\begin{table}[thbp]
  \footnotesize
  \centering
  \bgroup
  \def\arraystretch{1.3}
  \begin{tabular}{rcccccc}
    \hline \hline
           &  \multicolumn{6}{c}{Composition [\%]} \\ 
           \hline

   Process & Baseline & High-mass & High-$p_\mathrm{T}$ & \search & \search, & \control \\ [-0.5ex]
           &           &           &         &          & $m_{jj}>1$~TeV &            \\
   \hline 
   \dyzjj   & 94.2 $\pm$ 0.4 & 86.8 $\pm$ 1.6 &92.3 $\pm$ 0.4 & 93.4 $\pm$ 0.9 & 72.9 $\pm$ 2.1 & 95.4 $\pm$ 0.8 \\
   \ewkzjj  &  \hspace{11pt}1.5 $\pm<$0.1 & 10.6 $\pm$ 0.2 & \hspace{11pt}2.6 $\pm<$0.1 &   \hspace{11pt}4.8 $\pm<$0.1 & 26.1 $\pm$ 0.5 & \hspace{11pt}1.6 $\pm<$0.1\\
 Diboson   &  \hspace{11pt}1.6 $\pm<$0.1 &  \hspace{5pt}1.5 $\pm$ 0.7 & \hspace{11pt}2.0 $\pm$ 0.5 &   \hspace{11pt}1.0 $\pm$0.5 & \hspace{5pt}0.8 $\pm$ 0.4 & \hspace{11pt}1.8 $\pm$0.4 \\ 
\hdashline
     \ttbar &  \hspace{11pt}2.6 $\pm<$0.1 &  \hspace{5pt}1.1 $\pm$ 0.1 & \hspace{4pt}3.1 $\pm$ 0.1 &   \hspace{11pt}0.7 $\pm<$0.1 &\hspace{5pt}0.1 $\pm$ 0.1 & \hspace{5pt}1.2 $\pm$ 0.1 \\
 Single-$t$ &         $<$0.2 &         $<$0.2 &         $<$0.2 &         $<$0.1 &         $<$0.1 & $<$0.1 \\
   Multijet &         $<$0.3 &         $<$0.3 &         $<$0.3 &         $<$0.3 &         $<$0.3 &$<$0.3 \\
   \hline\hline
   \multirow{2}{*}{Total expected}  & 64800 & 2220 & 21900 & 11100 &  640 & 7120  \\
                                     & $\pm$ 130 $\pm$ 5220 & $\pm$ 20 $\pm$ 200 & $\pm$ 40 $\pm$ 1210 & $\pm$ 50 $\pm$ 520 &  $\pm$ 10 $\pm$ 40 & $\pm$ 30 $\pm$ 880 \\

   \hline
   Total observed & 67472 & 1471 & 22461  & 11630 & 490 & 6453  \\
  \hline \hline

  \end{tabular}
  \egroup
  \caption{Estimated composition (in percent) of the data samples selected in the six \zjj fiducial regions for the dielectron and dimuon channels combined, using the \ewkzjj
    sample from \POWHEG, and the \dyzjj sample from \SHERPA (normalised using NNLO predictions for the inclusive \Z cross-section calculated with \textsc{FEWZ}). 
    Uncertainties in the sample contributions are statistical only.
Also shown are the total expected yields and the total observed yields in each fiducial region.
Uncertainties in the total expected yields are statistical (first) and systematic (second), see Section~\ref{sec:syst} for details.
}
  \label{table:compositionsSP}
\end{table}

The expected composition of the selected data samples in the six \zjj fiducial regions is summarised in Table~\ref{table:compositionsSP}, averaging across the dielectron and dimuon channels
as these compositions in the two dilepton channels are in agreement within statistical uncertainties.
The numbers of selected events in data and expectations from total signal plus background estimates
are also given for each region. 
The largest discrepancy between observed and expected yields is seen in the \highmass region, and results from a mismodelling of the \mjj spectrum in the \dyzjj MC simulations used,
which is discussed below and accounted for in the assessment of systematic uncertanties in the measurement.

\subsection{Systematic uncertainties in the inclusive \zjj fiducial cross-sections} 
\label{sec:syst}

Experimental systematic uncertainties affect the determination of the $\mathcal{C}^f$ correction factor and the background estimates.
The dominant systematic uncertainty in the inclusive \zjj fiducial cross-sections arises from the calibration of the jet energy scale and resolution.
This uncertainty varies from around 4\% in the \search region to around 12\% in the \control region. The larger uncertainty in the \control region is due to the higher average jet multiplicity
(an average of 1.7 additional jets in addition to the leading and sub-leading jets) compared with the \search region (an average of 0.4 additional jets).
Other experimental systematic uncertainties arising from lepton efficiencies related to reconstruction, identification, isolation and trigger, and lepton energy/momentum scale and resolution as well as from the effect of pile-up, amount to a total of around 1--2\%, depending on the fiducial region.

The systematic uncertainty arising from the MC modelling of the \mjj distribution in the \dyzjj\ and \ewkzjj\ signal processes is around 3\% in the \search
region, around 1\% in the \control region, 2\% in the \highmass region, and below 1\% elsewhere.
This is assessed by comparing the
correction factors obtained by using the different MC event generators listed in Section~\ref{samples} and by performing a data-driven reweighting of the \dyzjj\ MC sample to describe the \mjj distribution of the observed data in a given fiducial region.
Additional contributions arise from varying the QCD renormalisation and
factorisation scales up and down by a factor of two independently, and from the propagation of uncertainties in the PDF sets.
The normalisation of the diboson contribution is varied according to corresponding to PDF and scale variations in these predictions\,\cite{ATL-PHYS-PUB-2016-002}, 
and results in up to a $0.1\%$ effect on the measured \zjj cross-sections depending on the fiducial region. 
The uncertainty from varying the normalisation and shape in $\mjj$ of the estimated background from
top-quark production is at most 1\% (in the \highmass region), arising 
from changes in the extracted \zjj cross-sections when using modified top-quark background MC samples with PDF and scale variations, suppressed or enhanced additional radiation (generated with the \textsc{Perugia2012radHi/Lo} tunes\,\cite{perugia}), or
using an  alternative top-quark production sample from \MGMCatNLO interfaced to \textsc{Herwig++} v2.7.1\,\cite{Alwall:2014hca,Bahr:2008pv}.

The systematic uncertainty in the integrated luminosity is 2.1\%. 
This is derived following a methodology similar to
that detailed in Ref.~\cite{DAPR-2013-01}, from a calibration of
the luminosity scale using $x$--$y$ beam-separation scans performed in June 2015.

\subsection{Inclusive \zjj results}
\label{sec:inclresults}

The measured cross-sections in the dielectron and dimuon channels are combined and presented here as a weighted average (taking into account total uncertainties) across both channels.
These cross-sections are determined using each of the correction factors derived from the six combinations of the three \dyzjj (\ALPGEN, \MADGRAPH, and \SHERPA) and two \ewkzjj (\POWHEG and \SHERPA) MC samples.
For a given fiducial region (Table~\ref{table:phasespacescuts}) 
the cross-section averaged over all six variations
is presented in Table~\ref{table:QCDEW_xsec}. 
The envelope of variation between \dyzjj and \ewkzjj models is assigned as a source of systematic uncertainty
(1\% in all regions except the \search region where the variation is 3\% and the \highmass region where the variation is 2\%).
\begin{table}[htbp]
  \footnotesize
  \centering
  \bgroup
  \def\arraystretch{1.3}
  \begin{tabular}{r|rlll|rl|rl|rl}
    \hline \hline
    \multirow{2}{*}{Fiducial region} &  \multicolumn{10}{c}{Inclusive \zjj cross-sections [pb]} \\
    \cline{2-11}
                                     &  \multicolumn{4}{c|}{\multirow{2}{*}{Measured}} &  \multicolumn{6}{c}{Prediction} \\
                                     &  \multicolumn{4}{c|}{value $\pm$ stat. $\pm$ syst. $\pm$ lumi.} & \multicolumn{2}{c|}{\SHERPA(\dyzjj)} &  \multicolumn{2}{c|}{\MADGRAPH(\dyzjj)} &  \multicolumn{2}{c}{\ALPGEN(\dyzjj)} \\ [-0.9ex] 
                                     &                                    &&&           &  \multicolumn{2}{c|}{+\POWHEG(\ewkzjj)} &  \multicolumn{2}{c|}{+\POWHEG(\ewkzjj)} &  \multicolumn{2}{c}{+\POWHEG(\ewkzjj)} \\ 

           \hline
 Baseline & 13.9\!\!\!\!\! & $\pm$ 0.1\!\!\!\!\! & $\pm$  1.1\!\!\!\!\! & $\pm$  0.3 & $13.5$\!\!\!\!\!  & $\pm$ 1.9 & $15.2$\!\!\!\!\!  & $\pm$ 2.2 & $11.7$\!\!\!\!\!  & $\pm$ 1.7 \\
 High-$p_\mathrm{T}$ & 4.77\!\!\!\!\! & $\pm$ 0.05\!\!\!\!\! & $\pm$ 0.27\!\!\!\!\! & $\pm$ 0.10 & $4.7$\!\!\!\!\!  & $\pm$ 0.8 & $5.5$\!\!\!\!\!  & $\pm$ 0.9 & $4.2$\!\!\!\!\!  & $\pm$ 0.7 \\
 \search & 2.77\!\!\!\!\! & $\pm$ 0.04\!\!\!\!\! & $\pm$ 0.13\!\!\!\!\! & $\pm$ 0.06 & $2.7$\!\!\!\!\!  & $\pm$ 0.2 & $3.6$\!\!\!\!\!  & $\pm$ 0.3 & $2.4$\!\!\!\!\!  & $\pm$ 0.2 \\
 QCD-enriched & 1.34\!\!\!\!\! & $\pm$ 0.02\!\!\!\!\! & $\pm$ 0.17\!\!\!\!\! & $\pm$ 0.03 & $1.5$\!\!\!\!\!  & $\pm$ 0.4 & $1.4$\!\!\!\!\!  & $\pm$ 0.3 & $1.1$\!\!\!\!\!  & $\pm$ 0.3 \\
 High-mass & 0.30\!\!\!\!\! & $\pm$ 0.01\!\!\!\!\! & $\pm$ 0.03\!\!\!\!\! & $\pm$ 0.01  & $0.46$\!\!\!\!\!  & $\pm$ 0.11 & $0.40$\!\!\!\!\!  & $\pm$ 0.09 & $0.27$\!\!\!\!\!  & $\pm$ 0.06 \\
 \search  & \multirow{2}{*}{0.118\!\!\!\!\!} & \multirow{2}{*}{$\pm$ 0.008\!\!\!\!\!} & \multirow{2}{*}{$\pm$ 0.006\!\!\!\!\!} & \multirow{2}{*}{$\pm$ 0.002}  
                                          & \multirow{2}{*}{$0.156$\!\!\!\!\!}  & \multirow{2}{*}{$\pm$ 0.019} & \multirow{2}{*}{$0.185$\!\!\!\!\!}  & \multirow{2}{*}{$\pm$ 0.023} & \multirow{2}{*}{$0.120$\!\!\!\!\!}  & \multirow{2}{*}{$\pm$ 0.015} \\ [-0.9ex]
 ($m_{jj}>1$~TeV) & & & & & & &&&\\
   \hline\hline
  \end{tabular}
  \egroup
  \caption{Measured and predicted inclusive \zjj production cross-sections in the six fiducial regions defined in
    Table~\ref{table:phasespacescuts}.
    For the measured cross-sections, the first uncertainty given is statistical, the second is systematic and the third is due to the
    luminosity determination. 
    For the predictions, the statistical uncertainty is added in quadrature to the systematic uncertainties arising from the PDFs and factorisation and renormalisation scale variations.
  }
  \label{table:QCDEW_xsec}
\end{table}

The theoretical predictions from 
\SHERPA(\dyzjj) + \POWHEG(\ewkzjj),
\MADGRAPH(\dyzjj) + \POWHEG(\ewkzjj), and
\ALPGEN(\dyzjj) + \POWHEG(\ewkzjj) 
are found to be in agreement with the measurements in most cases.
The uncertainties in the theoretical predictions are significantly larger than the
uncertainties in the corresponding measurements.

The largest differences between predictions and measurement are in the \highmass and \search ($\mjj>250$~GeV and $>1$~TeV) regions.
Predictions from \SHERPA(\dyzjj) + \POWHEG(\ewkzjj) and \MADGRAPH(\dyzjj) + \POWHEG(\ewkzjj) 
exceed measurements in the \highmass region by 
54\% and 34\% respectively, where the predictions have relative uncertainties with respect to the measurement of 36\% and 32\%.
For the \search region, \SHERPA(\dyzjj) + \POWHEG(\ewkzjj) describes the observed rates well, but  \MADGRAPH(\dyzjj) + \POWHEG(\ewkzjj) overestimates measurements by 28\% with a relative uncertainty of 11\%.
In the \search ($\mjj>1$~TeV) region the same predictions overestimate measured rates by 33\% and 57\%, with relative uncertainties of 16\% and 15\%.
Some of these 
differences arise from a significant mismodelling of the \dyzjj contribution, as investigated and discussed in detail in Section~\ref{sec:mjjmismodelling}.
Predictions from \ALPGEN(\dyzjj) + \POWHEG(\ewkzjj) are in agreement with the data for the \highmass and \search ($\mjj> 250$~GeV and $>1$~TeV) regions.

\section{Measurement of \ewkzjj fiducial cross-sections} 

The \search fiducial region (defined in Table~\ref{table:phasespacescuts}) is used to measure the production cross-section of the \ewkzjj process.
The \search region has an overall expected  \ewkzjj signal fraction of 4.8\%
(Table~\ref{table:compositionsSP}) and this signal fraction grows with increasing \mjj to 26.1\% for $\mjj>1$~TeV.
The \control region has an overall expected  \ewkzjj signal fraction of 1.6\% increasing to 4.4\% for $\mjj>1$~TeV.
The dominant background to the \ewkzjj cross-section measurement is \dyzjj production.
It is subtracted in the same way as non-\zjj backgrounds in the inclusive measurement described in Section~\ref{inclusive}.
Although diboson production includes contributions from purely EW processes, in this measurement it is considered as part of the background and is estimated from simulation.

A particle-level production cross-section measurement of \ewkzjj production in a given fiducial region $f$ is thus given by
\begin{equation}
  \sigma^{f}_\mathrm{EW} = \frac{N^f_\mathrm{obs} - N^f_{\mathrm{QCD}-\zjj} - N^f_\mathrm{bkg}}{L \cdot \mathcal{C}_\mathrm{EW}^f},
\label{eqn:ewk}
\end{equation}
with the same notations as in Eq.~(\ref{eqn:incl}) and where
$N^f_{\mathrm{QCD}-\zjj}$ is the expected number of \dyzjj events passing the selection requirements of the fiducial region at detector level,
$N^f_\mathrm{bkg}$ is the expected number of background (non-\zjj and diboson) events, 
and $\mathcal{C}_\mathrm{EW}^f$ is a correction factor applied to the observed background-subtracted 
data yields that accounts for experimental efficiency and detector resolution effects, and is derived from \ewkzjj MC simulation with data-driven efficiency and energy/momentum scale corrections.
For the $\mjj>250$~GeV ($\mjj>1$~TeV) region this correction factor is determined to be $0.66$ ($0.67$) when using the \SHERPA \ewkzjj prediction, and $0.67$ ($0.68$) when using the \POWHEG \ewkzjj prediction.

\begin{figure}[htbp]
  \centering
  \begin{subfigure}[b]{0.49\textwidth}
    \includegraphics[scale=0.4]{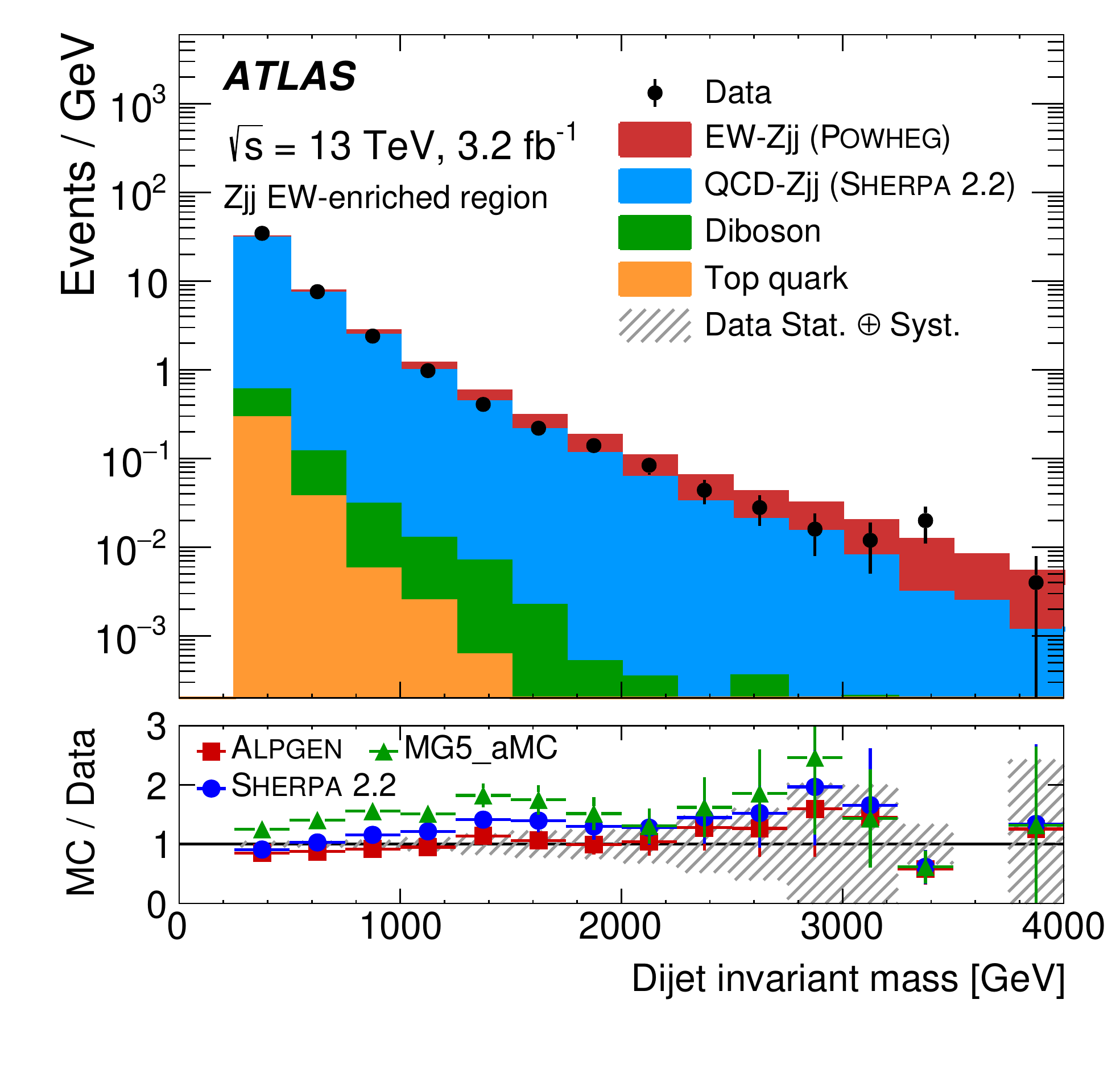}
    \caption{}\label{fig:mjjsearchSP}
  \end{subfigure}
  \begin{subfigure}[b]{0.49\textwidth}
    \includegraphics[scale=0.4]{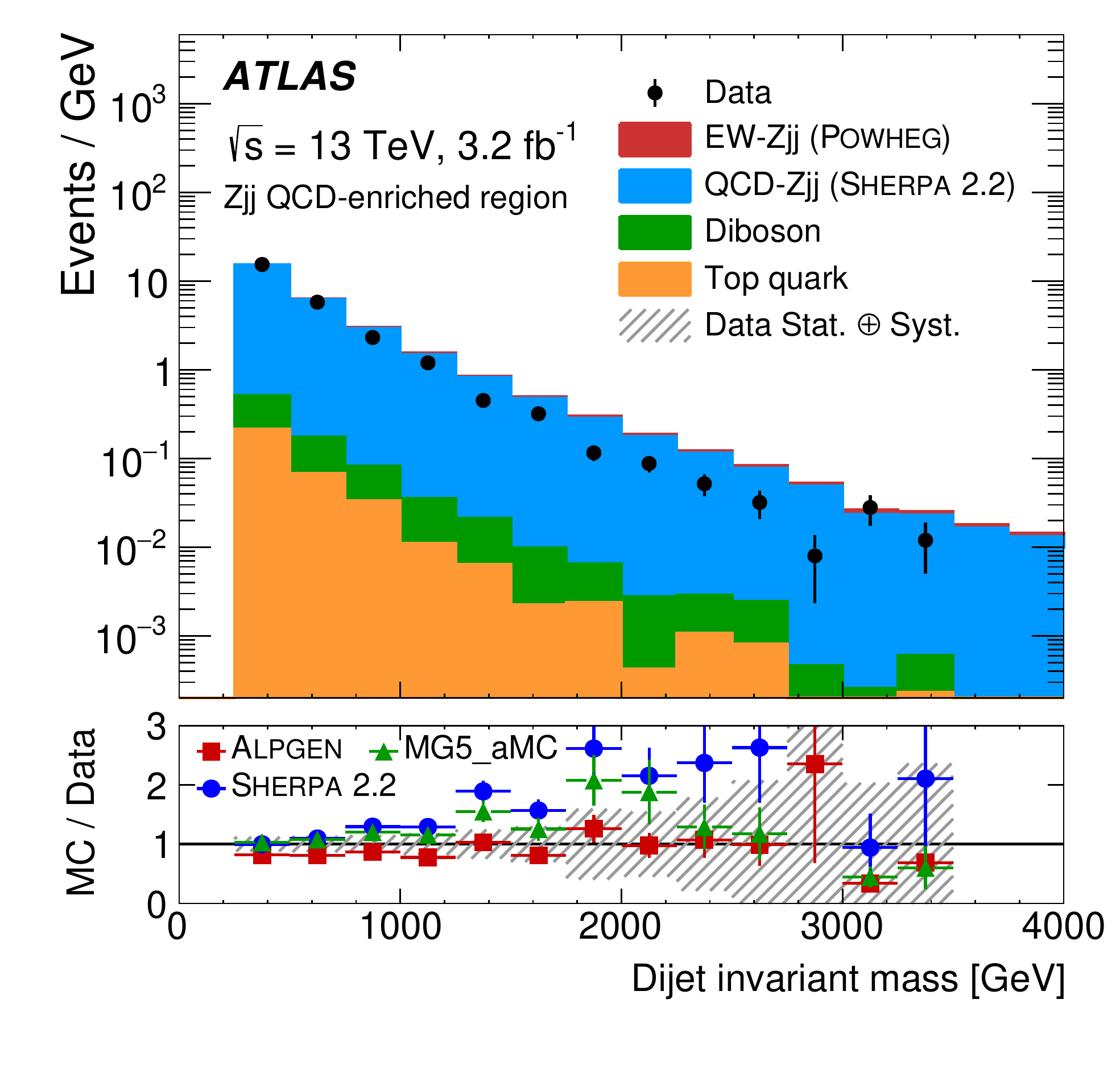}
    \caption{}\label{fig:mjjcontrolSP}
  \end{subfigure}
  \caption{Detector-level comparisons of the dijet invariant mass distribution between data and simulation in (a)~the \search region and
    (b)~the \control region, for the dielectron and dimuon channel combined. Uncertainties shown on the data are statistical only. 
    The \ewkzjj simulation sample comes from the \POWHEG event generator and the \dyzjj MC sample comes from the \SHERPA event generator. 
    The lower panels show the ratio of simulation to data for three \dyzjj models, from  \ALPGEN, \MADGRAPH, and \SHERPA.
    The hatched band centred at unity represents the size of statistical and experimental systematic uncertainties added in quadrature.}
\label{fig:M_j1j2-SP}
\end{figure}

Detector-level comparisons of the $m_{jj}$ distribution between data and simulation in (a)~the \search region and
(b)~the \control region are shown in Figure~\ref{fig:M_j1j2-SP}.
It can be seen in Figure~\ref{fig:M_j1j2-SP}(a) that in the \search region the \ewkzjj component becomes prominent at large values of \mjj.
However, Figure~\ref{fig:M_j1j2-SP}~(b) demonstrates that the shape of the  $m_{jj}$ distribution for \dyzjj production is poorly modelled in simulation.
The same trend is seen for all three \dyzjj\ event generators listed in Section~\ref{samples}. \ALPGEN provides the best description  of the data over the whole \mjj range. 
In comparison, \MADGRAPH and \SHERPA overestimate the data by 80\% and 120\% respectively, for \mjj = 2~TeV, well outside the uncertainties on these predictions described in Table~\ref{table:QCDEW_xsec}.
These discrepancies have been observed previously in \zjj\,\cite{8TeVVBFNote,STDM-2016-01} and \wjj\,\cite{Abazov:2013gpa,Khachatryan:2016fue,STDM-2014-11} production at high dijet invariant mass and at high jet rapidities.
For the purpose of extracting the cross-section for  \ewkzjj production, this mismodelling  of \dyzjj is corrected for
    using a data-driven approach, as discussed in the following.

\subsection{Corrections for mismodelling of \dyzjj production and fitting procedure} 
\label{sec:mjjmismodelling}

The normalisation of the \dyzjj background is extracted from a fit of the \dyzjj and \ewkzjj \mjj simulated distributions 
to the data in the \search region, after subtraction of non-\zjj and diboson background,
using a log-likelihood maximisation\,\cite{2003physics...6116V}.
Following the procedure adopted in Ref.\,\cite{8TeVVBFNote}, the data in the \control region are used to evaluate 
 detector-level shape correction factors for the  \dyzjj MC predictions bin-by-bin in \mjj.
These data-to-simulation ratio correction factors are applied to the simulation-predicted shape in \mjj of the \dyzjj contribution in the \search region.
This procedure is motivated by two observations:
\begin{enumerate}[label=(\alph*)]
\item the \control region and \search region are designed to be kinematically very similar, differing only with
regard to the presence/absence of jets reconstructed within the rapidity interval bounded by the dijet system,
\item the contribution of \ewkzjj to the region of high \mjj is suppressed in the
\control region (4.4\% for $\mjj>1$~TeV) relative to that in the \search region (26.1\% for $\mjj>1$~TeV) (also illustrated in Figure~\ref{fig:M_j1j2-SP}); 
the impact of the residual \ewkzjj contamination in the \control region is 
assigned as a component of the systematic uncertainty in the \dyzjj background.
\end{enumerate}

\begin{figure}[htbp]
  \centering
  \begin{subfigure}[b]{0.49\textwidth}
    \includegraphics[scale=0.39]{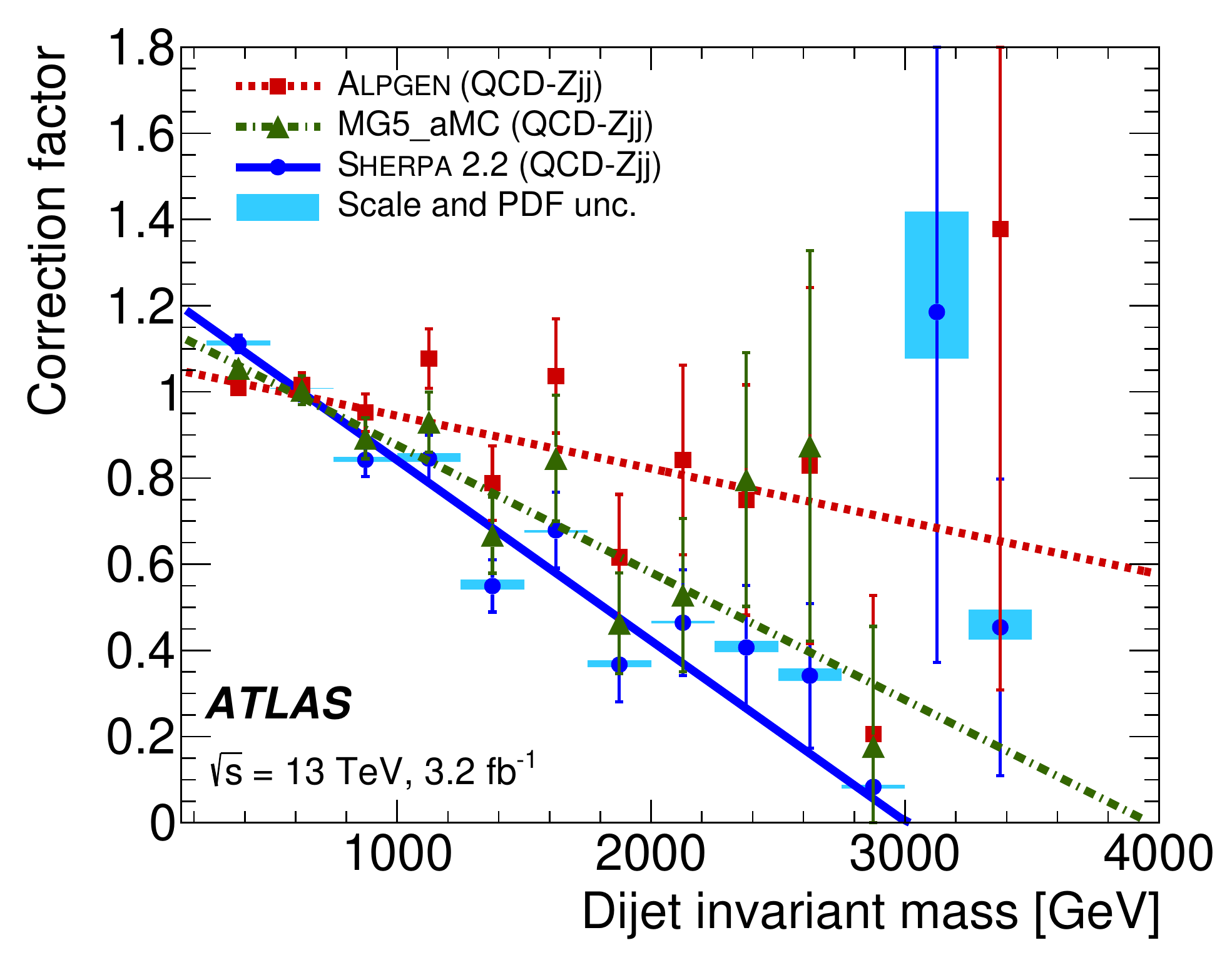}
    \caption{}
  \end{subfigure}
  \begin{subfigure}[b]{0.49\textwidth}
    \includegraphics[scale=0.39]{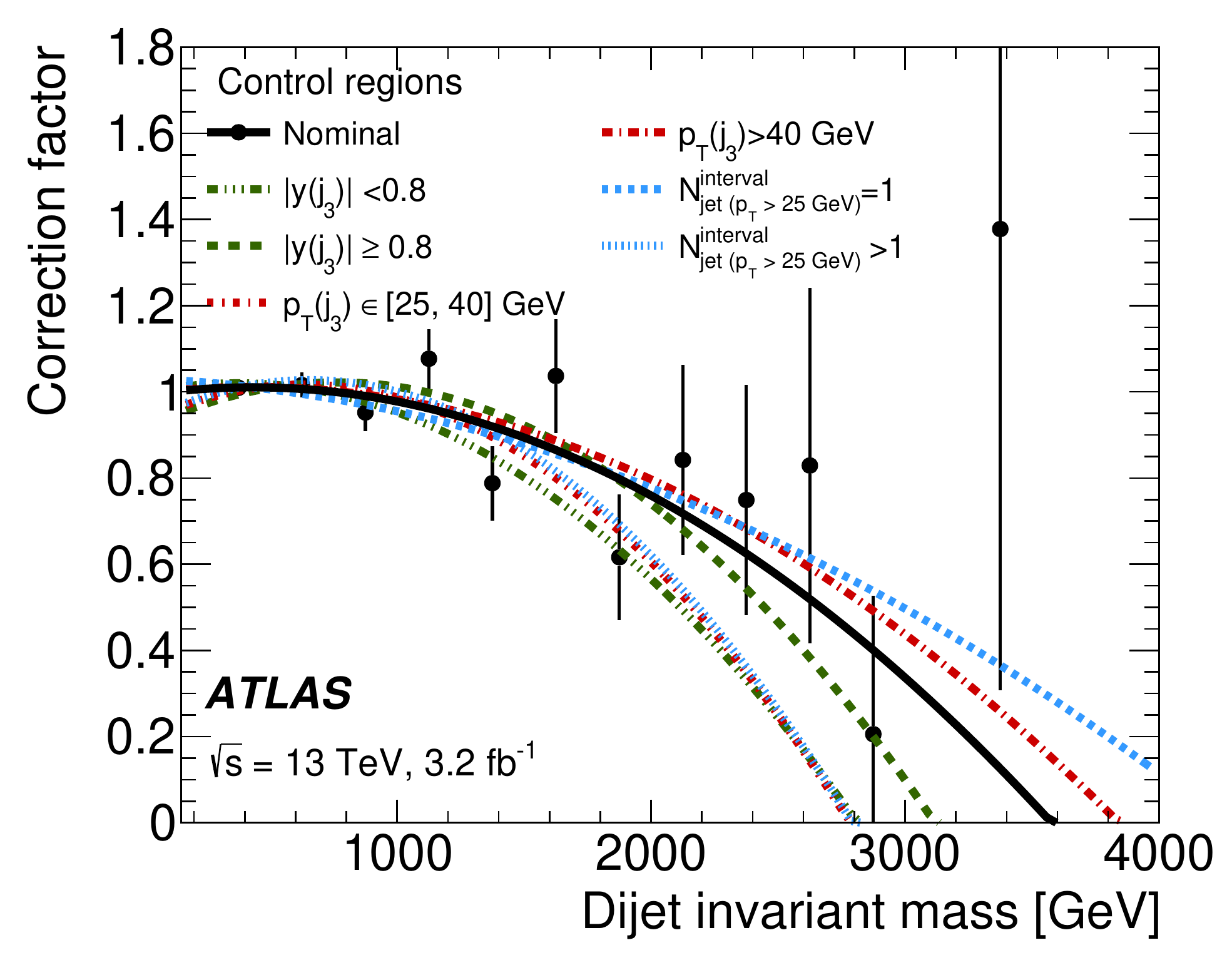}
    \caption{}
  \end{subfigure}
  \caption{
Binned data-to-simulation normalised ratio shape correction factors as a function of dijet invariant mass in the \control region.
 (a)~Ratio for three different \dyzjj MC samples with uncertainties corresponding to the combined statistical
uncertainties in the data and QCD-Zjj MC samples added in quadrature. 
 Scale and PDF uncertainties in \SHERPA predictions are indicated by the shaded bands. 
 Lines represent fits to the ratios using a linear fit.
 (b)~Ratio for subregions of the \control region for the \ALPGEN MC sample. Curves represent
the result of fits with a quadratic function for the various subregions.
}
  \label{fig:fit}
\end{figure}

The shape correction factors in \mjj obtained using the three different \dyzjj MC samples are shown in
Figure~\ref{fig:fit}~(a). These are derived as the ratio of the data to simulation in bins of \mjj after normalisation of the total yield in simulation to that observed in data in the \control region.
A binned fit to the correction factors derived in dijet invariant mass is performed with a linear fit function (and also with a quadratic fit function) to produce a continuous correction factor.
The linear fit is illustrated overlaid on the binned correction factors in Figure~\ref{fig:fit}~(a).
The nominal value of the \ewkzjj cross-section corresponding to a particular \dyzjj event generator template 
is determined using the correction factors from the linear fit.
The change in resultant \ewkzjj cross-section from using 
binned correction factors directly is assessed as a systematic uncertainty.
The change in the extracted \ewkzjj cross-section when using a quadratic fit was found to be negligible.
The variations observed between event generators 
may be partly due to differences 
in the modelling of QCD radiation  within the rapidity interval bounded by the dijet system, which affects the extrapolation from the central-jet-enriched \control region
to the central-jet-suppressed \search region. The variation between event generators is much larger than the effect of PDF and
scale uncertainties in a particular prediction (indicated in Figure~\ref{fig:fit}~(a) by a shaded band on the predictions from  \SHERPA).
Estimating the uncertainties associated with \dyzjj mismodelling from PDF and scale variations around a single generator prediction would thus result in an underestimate of the true theoretical uncertainty associated with this mismodelling.
In this measurement, the span of resultant \ewkzjj cross-sections extracted from the use of each of the three \dyzjj templates is assessed as a systematic uncertainty.
The variation in the \ewkzjj cross-section measurement due to a change in the \ewkzjj signal template used in the derivation of the
\mjj correction factors (from \POWHEG to \SHERPA) is found to be negligible.

\begin{figure}[htbp]
  \centering
  \begin{subfigure}[b]{0.49\textwidth}
    \includegraphics[scale=0.39]{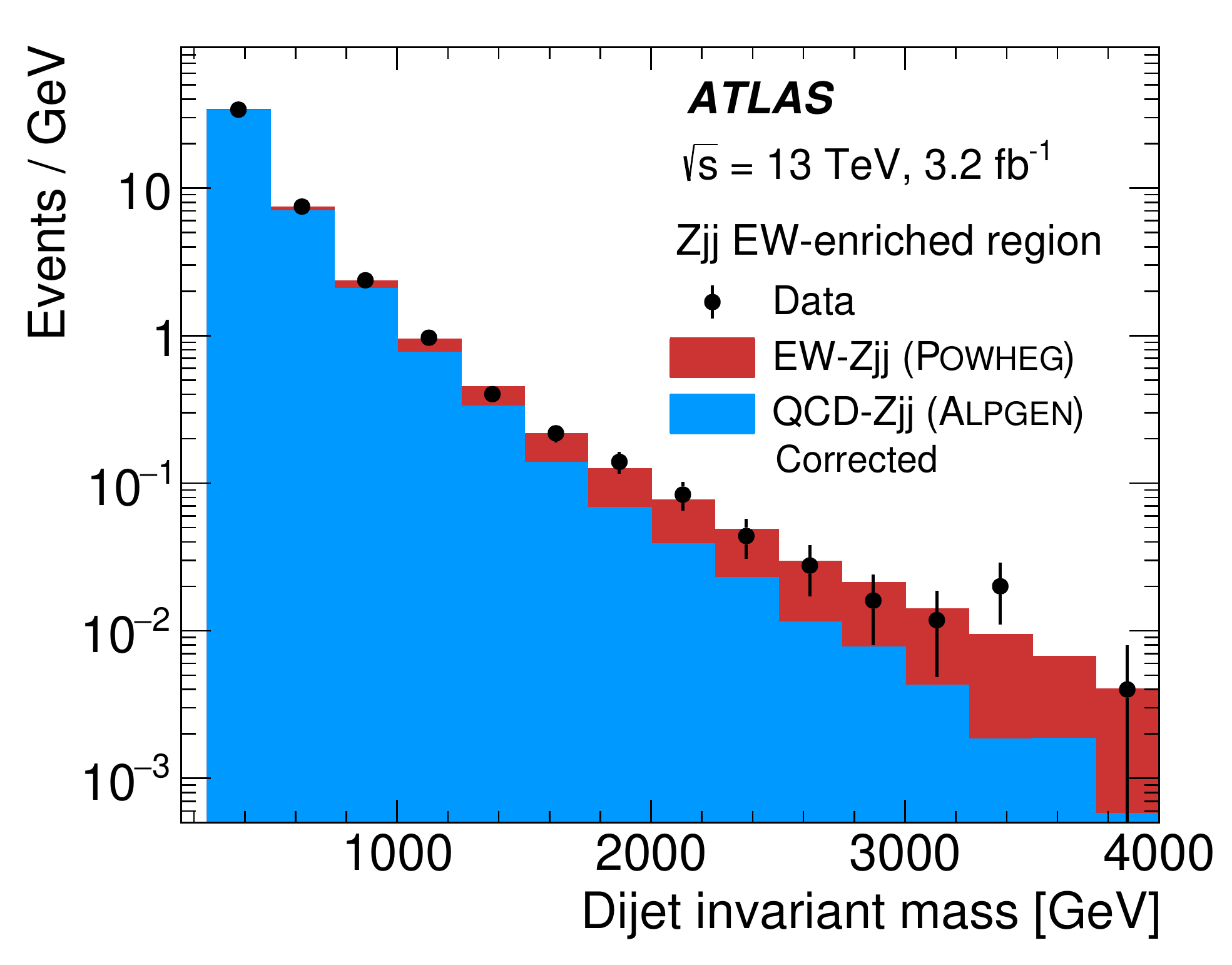}
    \caption{}
  \end{subfigure}
  \begin{subfigure}[b]{0.49\textwidth}
    \includegraphics[scale=0.39]{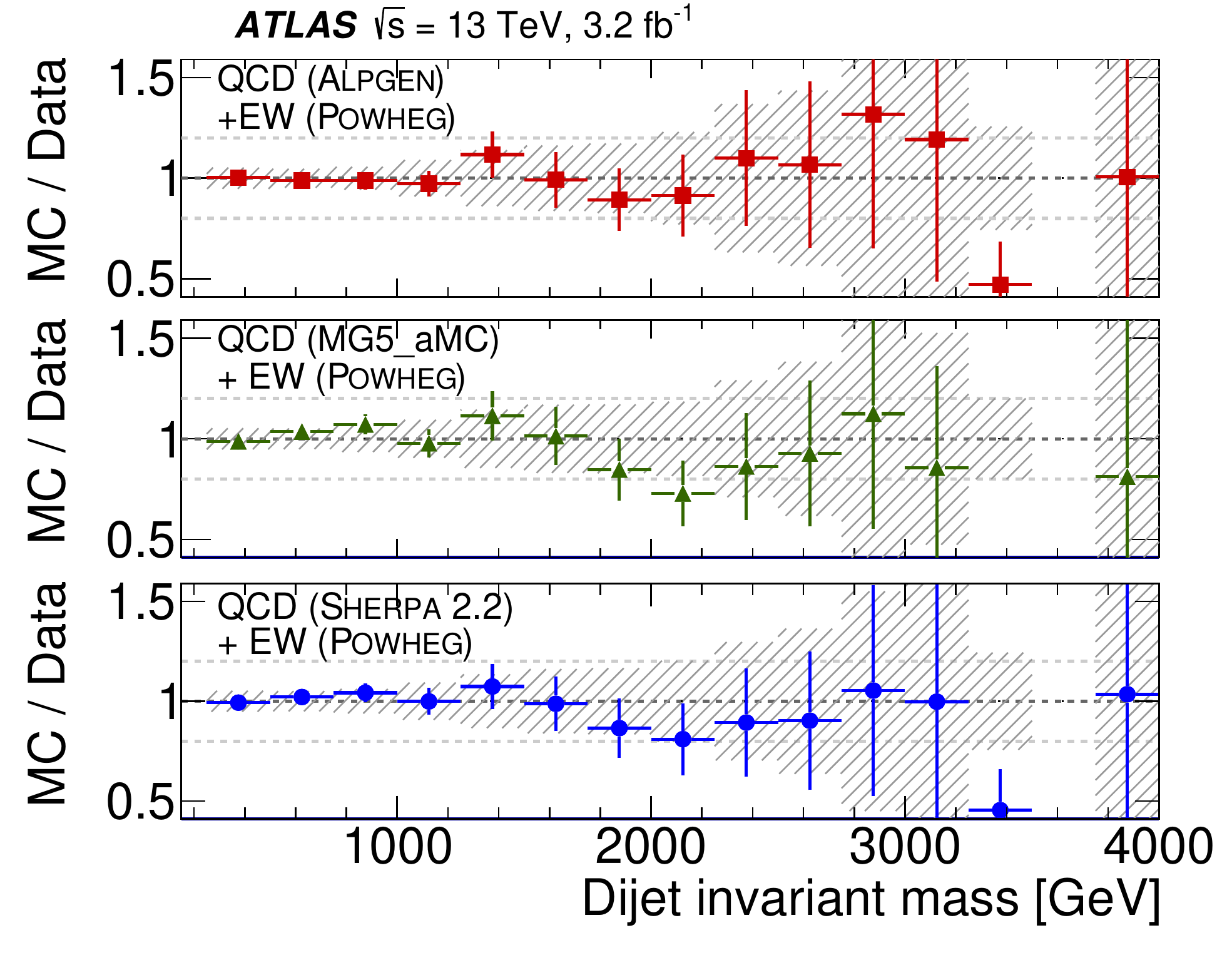}
    \caption{}
  \end{subfigure}
  \caption{
 (a)~Comparison in the \search region of the sum of \ewkzjj and \mjj-reweighted \dyzjj templates to the data (minus the non-\zjj
      backgrounds). 
      The normalisation of the templates is adjusted to the results of the fit (see text for details).
      The \ewkzjj MC sample comes from the \powheg event generator and the \dyzjj MC sample comes from the  \ALPGEN event generator.
 (b)~The ratio of the sum of the \ewkzjj and \mjj-reweighted \dyzjj templates to the background-subtracted data in the \search region, 
 for three different \dyzjj MC predictions.
      The normalisation of the templates is adjusted to the results of the fit. Error bars represent the statistical uncertainties in the data and combined \dyzjj plus \ewkzjj MC samples added in quadrature.
      The hatched band represents experimental systematic uncertainties in the \mjj distribution.
}
  \label{fig:fitresult}
\end{figure}

To test the dependence of the \dyzjj correction factors on the modelling of any additional jet emitted in the dijet rapidity interval, 
the \control control region is divided into pairs of mutually exclusive subsets according to
the $|y|$ of the highest \pT jet within the rapidity interval bounded by the dijet system, 
the \pT  of that jet,
or the value of \ngap.
The continuous correction factors are determined from each subregion 
using both a linear and a quadratic fit to the data.
Correction factors derived in the subregions using quadratic fits result in the largest variation in the extracted cross-sections. These fits are shown in Figure~\ref{fig:fit}~(b) for the \ALPGEN \dyzjj sample, 
which displays the largest variation between subregions of the three event generators used to produce \dyzjj predictions.
Within statistical uncertainties the measured \ewkzjj cross-sections are not sensitive to the definition of the control region used.

The normalisations of the corrected \dyzjj templates and the \ewkzjj templates are allowed to vary independently in a fit to the background-subtracted \mjj distribution in the \search region.
The measured electroweak production cross-section is determined from the data minus the \dyzjj contribution determined from these fits (Eq.~(\ref{eqn:ewk})).
As the choice of \ewkzjj template can influence the normalisation of the \dyzjj template in the \search region fit, the measured \ewkzjj cross-section determination is repeated for each \dyzjj template
using either the \POWHEG or \SHERPA \ewkzjj template in the fit. The central value of the result quoted is the
average
of the measured \ewkzjj cross-sections determined with each of the six combinations of the three \dyzjj and two \ewkzjj templates, with the envelope of measured results from these variations taken as an uncertainty associated with the dependence on the modelling of
the templates in the \search region. Separate uncertainties are assigned for the determination of the \dyzjj correction factors in the \control region and their propagation into the \search region.
The measurement of the \ewkzjj cross-section in the \search region for $\mjj>1$~TeV is extracted from the same fit procedure, with data and \dyzjj yields integrated for $\mjj>1$~TeV. 

Figure~\ref{fig:fitresult}~(a) shows a comparison in the \search region of the fitted \ewkzjj and \mjj-reweighted \dyzjj templates to the background-subtracted data, from which the measured \ewkzjj cross-section is extracted.
Figure~\ref{fig:fitresult}~(b) demonstrates how the data in the \search region is modelled with the fitted \ewkzjj and \mjj-reweighted \dyzjj templates, 
for the three different \dyzjj event generators (and their corresponding correction factors derived in the \control region
shown in Figure~\ref{fig:fit}~(a)). 
Despite significantly different modelling of the \mjj distribution between event generators, and different models for additional QCD radiation, the results of the combined correction and fit procedure give a consistent description of the data.

\subsection{Systematic uncertainties in the  \ewkzjj fiducial cross-section} 

The total systematic uncertainty in the cross-section for \ewkzjj production in the \search fiducial region is 17\% (16\% in the \search $\mjj>1$~TeV region).
The sources and size of each systematic uncertainty 
are summarised in Table~\ref{table:EWKsystunc}.

\begin{table}[htbp]
  \centering
  \bgroup
  \def\arraystretch{1.1}
  \begin{tabular}{lrr}
    \hline \hline
    &    \multicolumn{2}{c}{Relative systematic uncertainty [\%]} \\
    Source & $\sigma^{\mjj>250~\mathrm{GeV}}_\mathrm{EW}$ & $\sigma^{\mjj>1~\mathrm{TeV}}_\mathrm{EW}$ \\
    \hline

    \ewkzjj signal modelling (QCD scales, PDF and UEPS)                           &    $\pm$ 7.4 &  $\pm$ 1.7 \\
    \ewkzjj template statistical uncertainty           &    $\pm$ 0.5 &  $\pm$ 0.04 \\ \hdashline

    \ewkzjj contamination in \control region           &      $-0.1$ &     $-0.2$ \\
    \dyzjj modelling (\mjj shape constraint / third-jet veto)   &   $\pm$ 11 &  $\pm$ 11 \\
    Stat. uncertainty in QCD control region constraint &    $\pm$ 6.2 &   $\pm$ 6.4 \\ 
    \dyzjj signal modelling (QCD scales, PDF and UEPS)    &    $\pm$ 4.5 &   $\pm$ 6.5 \\
    \dyzjj template statistical uncertainty            &    $\pm$ 2.5 &   $\pm$ 3.5 \\ \hdashline

    QCD--EW interference                               &    $\pm$ 1.3 &   $\pm$ 1.5 \\ \hdashline

    $\bar{t}t$ and single-top background modelling     &    $\pm$ 1.0 &   $\pm$ 1.2 \\
    Diboson background modelling                      &   $\pm$ 0.1 &  $\pm$ 0.1 \\ \hdashline

    Jet energy resolution                              &    $\pm$ 2.3 &   $\pm$ 1.1 \\
    Jet energy scale                                   &  +5.3/--4.1 & +3.5/--4.2 \\
    Lepton identification, momentum scale, trigger, pile-up        &  +1.3/--2.5 & +3.2/--1.5 \\ 
    Luminosity                                         &    $\pm$ 2.1 &   $\pm$ 2.1 \\
    \hline
    Total                                              &   $\pm$ 17 &  $\pm$ 16 \\
    \hline \hline
  \end{tabular}
  \egroup
  \caption{Systematic uncertainties contributing to the measurement of the \ewkzjj cross-sections for \mjj~$>$~250~GeV and \mjj~$>$~1~TeV.
  Uncertainties are grouped into \ewkzjj signal modelling, \dyzjj background modelling, QCD-EW interference, non-\zjj backgrounds,
  and experimental sources.}
  \label{table:EWKsystunc}
\end{table} 

Systematic uncertainties associated with the \ewkzjj signal template used in the fit and \ewkzjj signal extraction are obtained from the variation in the measured cross-section
when using either of the individual \ewkzjj MC samples (\POWHEG and \SHERPA) compared to the average of the two, taken as the central value.
Uncertainties in the \ewkzjj templates due to variations of the QCD scales, of the PDFs, and of the UEPS model are also included
as are statistical uncertainties in the templates themselves.

Following the extraction of the \ewkzjj cross-section in the \search regions, the normalisations of the \ewkzjj MC samples are modified to agree with the measurements and the potential EW contamination of the \control region is recalculated,
which leads to a modification of the \dyzjj correction factors. The \ewkzjj cross-section measurement is repeated with these modified \dyzjj templates and the change in the resultant cross-sections is assigned as a systematic uncertainty
associated with the \ewkzjj contamination of the \control region. 

As discussed in Section~\ref{sec:mjjmismodelling}, the use of a \control region provides a way to correct for \dyzjj modelling issues and also constrains 
theoretical and experimental uncertainties associated with observables constructed from the two leading jets. Nevertheless, 
the largest contribution to the total uncertainty arises from modelling uncertainties associated with propagation of the $m_{jj}$ correction factors for \dyzjj in the \control region into the \search region, 
and these correction factors depend on the modelling
of the additional jet activity in the \dyzjj MC samples used in the measurement. The uncertainty
is assessed by repeating the \ewkzjj cross-section measurement with \mjj-reweighted \dyzjj MC templates from \ALPGEN, \MADGRAPH, and \SHERPA, 
and assigning the variation of the measured cross-sections from the central \ewkzjj result as a systematic uncertainty.
Statistical uncertainties from data and simulation in the $m_{jj}$ correction factors derived in the \control region are also propagated through to the measured \ewkzjj cross-section as a systematic uncertainty.
Uncertainties associated with QCD renormalisation and factorisation scales, PDF error sets, and UEPS modelling are assessed by studying the change in the extracted \ewkzjj cross-sections when 
repeating the measurement procedure, including rederiving $m_{jj}$ correction factors in the \control region and repeating fits in the \search region, using modified \dyzjj MC templates.
Statistical uncertainties in the \dyzjj template in the \search region are also propagated as a systematic uncertainty in the \ewkzjj cross-section measurement.

Potential quantum-mechanical interference between the \dyzjj and \ewkzjj processes is assessed using \MADGRAPH to derive a correction to the \dyzjj template as a function of \mjj.
The impact of interference on the measurement is determined by repeating the \ewkzjj measurement procedure twice,
either applying this correction to the \dyzjj template only in the \control region or only in the \search region and taking the maximum change in the measured \ewkzjj cross-section as a symmetrised uncertainty.
This approach assumes the interference affects only one of the two fiducial regions and therefore has a maximal impact on the signal extraction.
Potential interference between the \zjj and diboson processes was found to be negligible.

Normalisation and shape uncertainties in the estimated background from top-quark and diboson production are assessed with varied background templates as described in Section~\ref{sec:syst},
albeit with significantly larger uncertainties in the \search fiducial region compared to the \baseline region.

Experimental systematic uncertainties arising from the jet energy scale and resolution, from lepton efficiencies related to reconstruction, identification, isolation and trigger, and lepton energy/momentum scale and resolution, and from pile-up modelling, are independently 
assessed by repeating the \ewkzjj measurement procedure using modified \dyzjj and \ewkzjj templates.
Here, the \control \dyzjj template constraint procedure described in Section~\ref{sec:mjjmismodelling}
has the added benefit of significantly reducing the jet-based experimental uncertainties, as can be seen in Table~\ref{table:EWKsystunc} from their small impact on the total systematic uncertainty.

\subsection{Electroweak \zjj results}
\label{sec:ewresults}

\begin{table}[htbp]
  \centering
  \bgroup
  \def\arraystretch{1.3}
  \begin{tabular}{l|llll|rl}
    \hline \hline
    \multirow{2}{*}{Fiducial region} &  \multicolumn{6}{c}{\ewkzjj cross-sections [fb]} \\
    \cline{2-7}
                                     &  \multicolumn{4}{c|}{Measured} &  \multicolumn{2}{c}{\powhegpythia} \\ 
           \hline
           \search, $m_{jj}>250$~GeV & \ewknum \!\!\!\!\!\! & $\pm$  \ewkstat   \!\!\!\!\!\! & $\pm$  \ewksyst   \!\!\!\!\!\! & $\pm$  \ewklumi  & $125.2$  \!\!\!\!\!\! & $\pm$ 3.4 \\ 
           \search, $m_{jj}>1$~TeV & \ewknumHM \!\!\!\!\!\! & $\pm$  \ewkstatHM \!\!\!\!\!\! & $\pm$  \ewksystHM \!\!\!\!\!\! & $\pm$  \ewklumiHM & $38.5$  \!\!\!\!\!\! & $\pm$ 1.5 \\ 
   \hline\hline
  \end{tabular}
  \egroup
  \caption{Measured and predicted \ewkzjj production cross-sections in the \search fiducial regions with and without an additional
    kinematic requirement of $m_{jj}>1$~TeV.
      For the measured cross-sections, the first uncertainty given is statistical, the second is systematic and the third is due to the
    luminosity determination. For the predictions, the quoted uncertainty represents the statistical uncertainty, plus systematic uncertainties from the PDFs and factorisation and renormalisation scale variations, all added in quadrature.
}
  \label{table:EW_xsec}
\end{table}

As in the inclusive \zjj cross-section measurements, the quoted \ewkzjj cross-section measurements are the average of the cross-sections determined with each of the six combinations of the three \dyzjj MC templates and two \ewkzjj MC templates.
The measured cross-sections for the EW production of a leptonically decaying \Z boson and at least two jets satisfying the fiducial
requirements for the \search regions as given in Table~\ref{table:phasespacescuts} with the requirements $\mjj>250$~GeV and $\mjj>1$~TeV
are shown in Table~\ref{table:EW_xsec}, where they are compared to predictions from \powhegpythia.
The use of a differential template fit in \mjj to extract the \ewkzjj signal allows systematic uncertainties on the \ewkzjj cross-section measurements to be constrained by the bins with the most favourable balance of \ewkzjj
signal purity and minimal shape and normalisation uncertainty. For the $\mjj>250$~GeV region, although all \mjj bins contribute to the fit, the individually most-constraining \mjj interval is the $900$--$1000$~GeV bin.
The use of this method results in very similar relative systematic uncertainties in the \ewkzjj cross-section measurements at the two different \mjj thresholds, despite the measured relative \ewkzjj contribution to the total \zjj rate 
for $\mjj>1$~TeV being more than six times the relative contribution of \ewkzjj for $\mjj>250$~GeV.

\begin{figure}[htbp]
  \centering
  \includegraphics[scale=0.67]{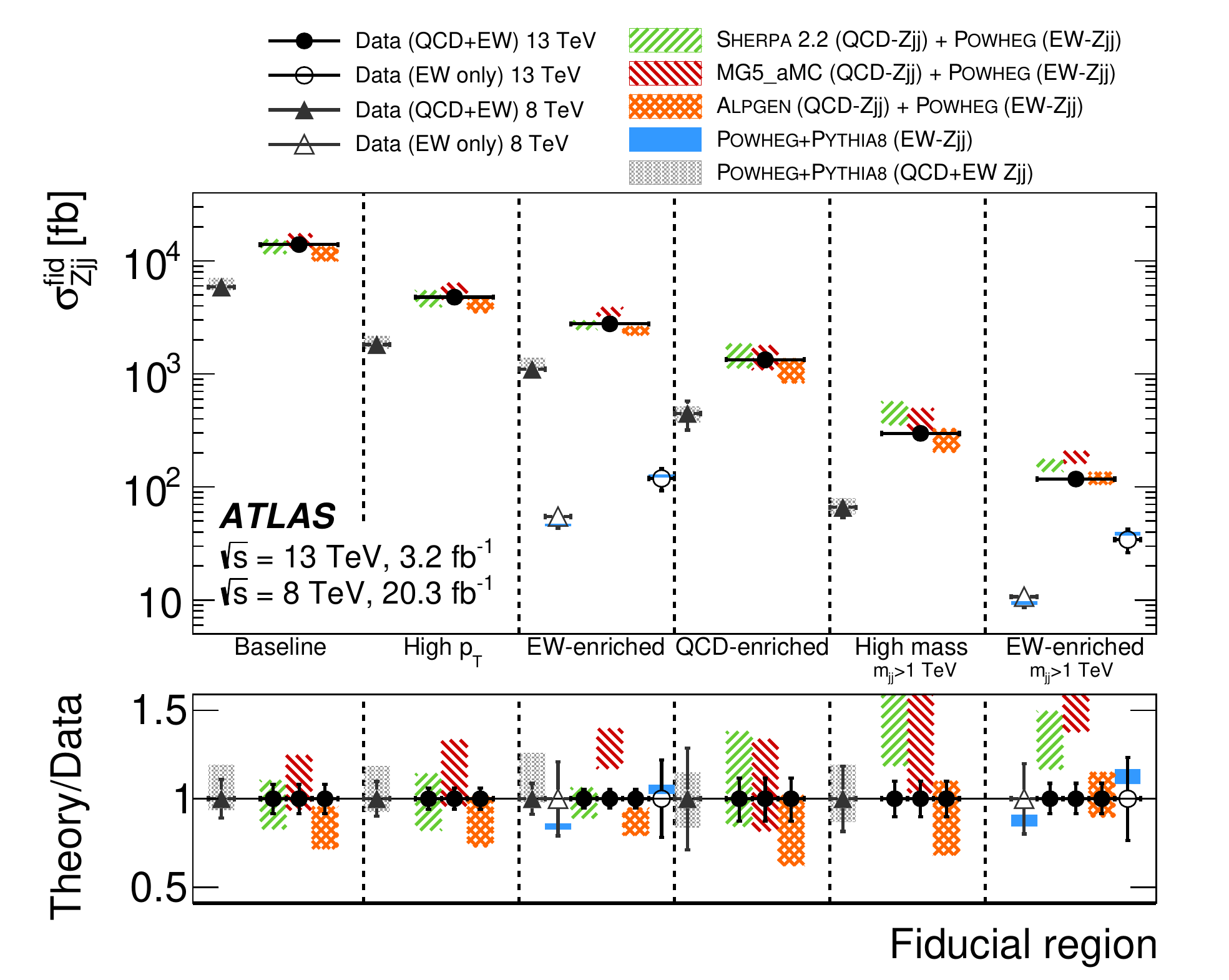}
\caption{Fiducial cross-sections for a leptonically decaying \Z boson and at least two jets 
(solid data points) and \ewkzjj production (open data points) at 13~TeV (circles) compared to equivalent results at 8~TeV\,\cite{8TeVVBFNote} (triangles) and to
theoretical predictions (shaded/hatched bands). Measurements of \zjj at 13~TeV are compared to predictions from \SHERPA(\dyzjj) + \POWHEG(\ewkzjj), \MADGRAPH (\dyzjj) + \POWHEG(\ewkzjj),  and \ALPGEN(\dyzjj) + \POWHEG(\ewkzjj),
while measurements of \ewkzjj production are compared to \POWHEG(\ewkzjj). Results at 8~TeV are compared to predictions from \powhegpythia(\dyzjj + \ewkzjj). 
The bottom panel shows the ratio of the various theory predictions to data as shaded bands. Relative uncertainties in the measured data are represented by an error bar centred at unity.}
  \label{fig:resultsSummary}
\end{figure}

The \ewkzjj cross-sections at  $\sqrt{s}=\SI{13}{\TeV}$ are in agreement with
the predictions from \powhegpythia for both \mjj~$>$~250~GeV and \mjj~$>$~1~TeV. 
The effect on the measurement of inclusive \zjj production rates (Section~\ref{sec:inclresults}) from correcting the \ewkzjj production rates predicted by \powhegpythia to the measured rates presented here was found to be negligible.
Modifications to the $\mjj$ distribution shape are already accounted for as a systematic uncertainty in the inclusive \zjj measurements.

\begin{figure}[htbp]
  \centering
  \includegraphics[scale=0.57]{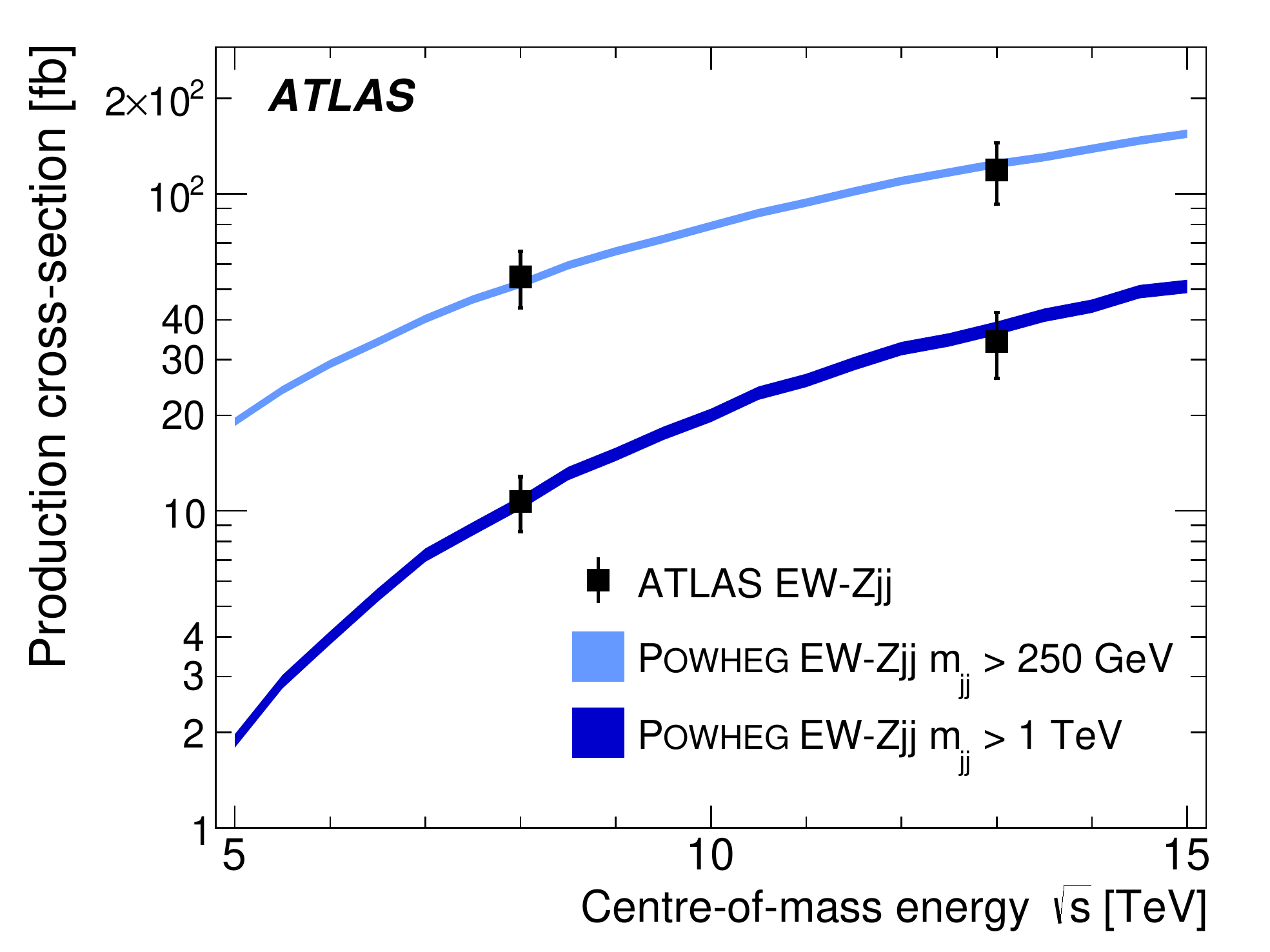}
  \caption{Measurements of the \ewkzjj process presented in this Letter at a centre-of-mass energy of 13~TeV, compared with previous measurements at 8~TeV\,\cite{8TeVVBFNote}, for two different dijet invariant mass thresholds, $m_{jj}>0.25$~TeV and $m_{jj}>1$~TeV. The error bars on the measurements represent statistical and systematic uncertainties added in quadrature. Predictions
  from the \POWHEG event generator with their total uncertainty are also shown.}
  \label{fig:sqrtS}
\end{figure}

Figure~\ref{fig:resultsSummary} shows a summary of the fiducial cross-sections for a leptonically decaying \Z boson and at least two jets 
at 13~TeV compared to equivalent results at 8~TeV\,\cite{8TeVVBFNote} and to theoretical predictions with their uncertainties.
A significant rise in cross-section is observed between $\sqrt{s}=\SI{8}{\TeV}$ and   $\sqrt{s}=\SI{13}{\TeV}$ within
each fiducial region. 
In the \search region, for \mjj thresholds of 250 GeV and 1~TeV, the measured \ewkzjj cross-sections at 13~TeV are 
found to be respectively 2.2 and 3.2 times as large as those measured at 8~TeV, as illustrated in Figure~\ref{fig:sqrtS}.

\FloatBarrier

\section{Summary}

    Fiducial cross-sections for the electroweak production of two 
    jets in association with a leptonically decaying \Z boson in proton--proton collisions are measured 
at a centre-of-mass energy of 13~TeV, using 
data corresponding to an integrated luminosity of 3.2~\ifb\ 
 recorded with the ATLAS detector at the Large Hadron Collider.
    The \ewkzjj  cross-section is extracted in a fiducial region chosen to enhance the EW contribution relative to the dominant \dyzjj process, which is constrained
    using a data-driven approach. The measured fiducial EW cross-section is  \sigmaewk~=~\ewkxsec~fb for dijet invariant mass greater than 250~GeV, and 
    $\ewkxsecHM$~fb for dijet invariant mass greater than 1~TeV.
 A comparison with previously published measurements at $\sqrt{s}=\SI{8}{\TeV}$ is presented, with measured \ewkzjj cross-sections at $\sqrt{s}=\SI{13}{\TeV}$ found to be 2.2 (3.2) times as large as those measured at $\sqrt{s}=\SI{8}{\TeV}$
in the low (high) dijet mass \search regions.
Relative to measurements at $\sqrt{s}=\SI{8}{\TeV}$, the increased $\sqrt{s}$ allows a region of higher dijet mass to be explored, in which the \ewkzjj signal is
more prominent.
The SM predictions are in agreement with the \ewkzjj measurements.

 The inclusive \zjj cross-section is also measured in six different fiducial regions with varying contributions from \ewkzjj and
 \dyzjj production. At higher dijet invariant masses ($>1$~TeV), particularly crucial for precision measurements of \ewkzjj production and for searches for new phenomena in vector-boson fusion topologies, 
 predictions from \SHERPA(\dyzjj) + \POWHEG(\ewkzjj) and \MADGRAPH (\dyzjj) + \POWHEG(\ewkzjj) are found to significantly overestimate the observed \zjj production rates in data.
 \ALPGEN(\dyzjj) + \POWHEG(\ewkzjj) provides a better description of the \mjj shape.

\section*{Acknowledgements}


We thank CERN for the very successful operation of the LHC, as well as the
support staff from our institutions without whom ATLAS could not be
operated efficiently.

We acknowledge the support of ANPCyT, Argentina; YerPhI, Armenia; ARC, Australia; BMWFW and FWF, Austria; ANAS, Azerbaijan; SSTC, Belarus; CNPq and FAPESP, Brazil; NSERC, NRC and CFI, Canada; CERN; CONICYT, Chile; CAS, MOST and NSFC, China; COLCIENCIAS, Colombia; MSMT CR, MPO CR and VSC CR, Czech Republic; DNRF and DNSRC, Denmark; IN2P3-CNRS, CEA-DRF/IRFU, France; SRNSFG, Georgia; BMBF, HGF, and MPG, Germany; GSRT, Greece; RGC, Hong Kong SAR, China; ISF, I-CORE and Benoziyo Center, Israel; INFN, Italy; MEXT and JSPS, Japan; CNRST, Morocco; NWO, Netherlands; RCN, Norway; MNiSW and NCN, Poland; FCT, Portugal; MNE/IFA, Romania; MES of Russia and NRC KI, Russian Federation; JINR; MESTD, Serbia; MSSR, Slovakia; ARRS and MIZ\v{S}, Slovenia; DST/NRF, South Africa; MINECO, Spain; SRC and Wallenberg Foundation, Sweden; SERI, SNSF and Cantons of Bern and Geneva, Switzerland; MOST, Taiwan; TAEK, Turkey; STFC, United Kingdom; DOE and NSF, United States of America. In addition, individual groups and members have received support from BCKDF, the Canada Council, CANARIE, CRC, Compute Canada, FQRNT, and the Ontario Innovation Trust, Canada; EPLANET, ERC, ERDF, FP7, Horizon 2020 and Marie Sk{\l}odowska-Curie Actions, European Union; Investissements d'Avenir Labex and Idex, ANR, R{\'e}gion Auvergne and Fondation Partager le Savoir, France; DFG and AvH Foundation, Germany; Herakleitos, Thales and Aristeia programmes co-financed by EU-ESF and the Greek NSRF; BSF, GIF and Minerva, Israel; BRF, Norway; CERCA Programme Generalitat de Catalunya, Generalitat Valenciana, Spain; the Royal Society and Leverhulme Trust, United Kingdom.

The crucial computing support from all WLCG partners is acknowledged gratefully, in particular from CERN, the ATLAS Tier-1 facilities at TRIUMF (Canada), NDGF (Denmark, Norway, Sweden), CC-IN2P3 (France), KIT/GridKA (Germany), INFN-CNAF (Italy), NL-T1 (Netherlands), PIC (Spain), ASGC (Taiwan), RAL (UK) and BNL (USA), the Tier-2 facilities worldwide and large non-WLCG resource providers. Major contributors of computing resources are listed in Ref.~\cite{ATL-GEN-PUB-2016-002}.

\printbibliography

\newpage 
\begin{flushleft}
{\Large The ATLAS Collaboration}

\bigskip

M.~Aaboud$^\textrm{\scriptsize 137d}$,
G.~Aad$^\textrm{\scriptsize 88}$,
B.~Abbott$^\textrm{\scriptsize 115}$,
O.~Abdinov$^\textrm{\scriptsize 12}$$^{,*}$,
B.~Abeloos$^\textrm{\scriptsize 119}$,
S.H.~Abidi$^\textrm{\scriptsize 161}$,
O.S.~AbouZeid$^\textrm{\scriptsize 139}$,
N.L.~Abraham$^\textrm{\scriptsize 151}$,
H.~Abramowicz$^\textrm{\scriptsize 155}$,
H.~Abreu$^\textrm{\scriptsize 154}$
% ...
\bigskip
\\
% ...
$^{12}$ Institute of Physics, Azerbaijan Academy of Sciences, Baku, Azerbaijan\\
% ...
$^{88}$ CPPM, Aix-Marseille Universit{\'e} and CNRS/IN2P3, Marseille, France\\
% ...
$^{115}$ Homer L. Dodge Department of Physics and Astronomy, University of Oklahoma, Norman OK, United States of America\\
% ...
$^{119}$ LAL, Univ. Paris-Sud, CNRS/IN2P3, Universit{\'e} Paris-Saclay, Orsay, France\\
% ...
$^{137}$ $^{(a)}$ Facult{\'e} des Sciences Ain Chock, R{\'e}seau Universitaire de Physique des Hautes Energies - Universit{\'e} Hassan II, Casablanca; $^{(b)}$ Centre National de l'Energie des Sciences Techniques Nucleaires, Rabat; $^{(c)}$ Facult{\'e} des Sciences Semlalia, Universit{\'e} Cadi Ayyad, LPHEA-Marrakech; $^{(d)}$ Facult{\'e} des Sciences, Universit{\'e} Mohamed Premier and LPTPM, Oujda; $^{(e)}$ Facult{\'e} des sciences, Universit{\'e} Mohammed V, Rabat, Morocco\\
% ...
$^{139}$ Santa Cruz Institute for Particle Physics, University of California Santa Cruz, Santa Cruz CA, United States of America\\
% ...
$^{151}$ Department of Physics and Astronomy, University of Sussex, Brighton, United Kingdom\\
% ...
$^{154}$ Department of Physics, Technion: Israel Institute of Technology, Haifa, Israel\\
$^{155}$ Raymond and Beverly Sackler School of Physics and Astronomy, Tel Aviv University, Tel Aviv, Israel\\
% ...
$^{161}$ Department of Physics, University of Toronto, Toronto ON, Canada\\
% ...
$^{a}$ Also at Department of Physics, King's College London, London, United Kingdom\\
$^{b}$ Also at Institute of Physics, Azerbaijan Academy of Sciences, Baku, Azerbaijan\\
$^{c}$ Also at Novosibirsk State University, Novosibirsk, Russia\\
$^{d}$ Also at TRIUMF, Vancouver BC, Canada\\
$^{e}$ Also at Department of Physics {\&} Astronomy, University of Louisville, Louisville, KY, United States of America\\
% ...
$^{*}$ Deceased
\end{flushleft}


\end{document}